%% file: main.tex
\documentclass[aps,prresearch,twocolumn,superscriptaddress,nofootinbib]{revtex4-2}

\usepackage{amsmath}
\usepackage{amssymb}
\usepackage{graphicx}
\usepackage{booktabs}
\usepackage{quantikz}
\usepackage{hyperref}

\usepackage[markup=underlined,addedmarkup=colored,deletedmarkup=sout]{changes}
\definechangesauthor[name={Ben}, color=blue]{BN}
\setdeletedmarkup{\textcolor{red}{\sout{#1}}}

\graphicspath{{figures/}}

\newtheorem{lemma}{Lemma}
\newenvironment{lemproof}{\par\noindent\emph{Proof.}\ }{\hfill$\blacksquare$\par\medskip}

\newcommand{\diag}{\operatorname{diag}}
\newcommand{\lamax}{\lambda_{\max}}
\newcommand{\Ht}{\widetilde{H}_1}
\newcommand{\At}{\widetilde{A}}

\begin{document}

\title{Convolution absorbing boundaries for explicit-circuit quantum
simulation of the wave equation}

\author{Hoang Anh Nguyen}
\email{hoanganh\_nguyen@mines.edu}
\affiliation{Department of Geophysics, Colorado School of Mines,
Golden, Colorado 80401, USA}

\author{Ali Tura}
\affiliation{Department of Geophysics, Colorado School of Mines,
Golden, Colorado 80401, USA}

\date{\today}

\begin{abstract}
Explicit quantum circuits for the wave equation, built by Hamiltonian
simulation, are restricted to closed domains, in which outgoing waves
reflect off the edge of the computational region and return.  We lift that
restriction with absorbing boundaries of the convolution type---a
complex-frequency-shifted perfectly matched layer realised through
exponential-kernel memory variables---and make the resulting non-unitary
dynamics quantum-implementable by Schr\"odingerisation.  We obtain three
results.  First, explicit gate-level circuits for the absorbing evolution,
obtained by extending a Bell-basis term-evolution circuit to
projector-valued operator strings and Trotterising to second order; these
run end to end on seventeen to twenty-one qubits and agree with exact
references to within a tenth of a percent to a percent.  Second, a
structural obstruction: the memory form of the absorbing generator carries
an irreducibly indefinite Hermitian part, whose largest eigenvalue grows
as the square root of the absorption strength divided by the grid spacing
and survives any diagonal rescaling of the memory fields.  The known
recovery threshold for Schr\"odingerisation then makes the post-selection
cost grow exponentially in the simulated time.  Third, a remedy: a
Lyapunov symmetrizer, precomputed classically, renders the transformed
generator dissipative and replaces that time-dependent penalty with a
time-independent conditioning factor of several hundred, measured on grids
of up to four thousand unknowns and saturating under mesh refinement.  The
crossover is early: past it the symmetrized recovery is cheaper by four to
twenty-six orders of magnitude in post-selection cost, and at the longest
horizons tested it is the only recovery that works.
\end{abstract}

\maketitle

\section{Introduction}
\label{sec:intro}

Quantum algorithms for partial differential equations (PDEs) divide into
oracle-based\footnote{oracle is an unitary black box, you can run but do not need to know what is inside} constructions, which assume block-encoded or sparse-access
operators \cite{costa2019wave,novikau2023qsvt}, and the explicit-circuit
line initiated for hyperbolic problems by Sato \emph{et
al.}~\cite{sato2024hamiltonian}, in which every gate of the time-stepping
circuit is written down and the cost is dominated by a number of
elementary operations polynomial in the register size.  Explicit-circuit constructions for wave-type equations published to date
operate on \emph{closed} domains: Dirichlet, Neumann, or periodic
boundaries, for which the semi-discrete generator is skew-Hermitian and
the evolution is genuinely
unitary~\cite{sato2024hamiltonian,bosch2025sources}; the extension to
nonconservative systems~\cite{sato2025nonconservative} admits a bulk
damping term $\zeta(x)\,\partial_t u$ via a linear combination of
Hamiltonian simulations, but its boundaries remain closed walls and its
demonstrations are confined to the undamped acoustic and uniform heat
equations.  Scattering and radiation problems,
however, are posed on open domains; truncating the domain with hard walls
reflects every outgoing wave back into the region of interest.  The same
requirement propagates into wave-based inverse problems: full waveform
inversion repeats the forward solve on a truncated domain at every
iteration, and hybrid quantum-classical workflows for it---in which
parameterized quantum circuits are embedded inside a domain-decomposed
physics-informed network and trained end to
end~\cite{nguyen2026hybrid}---inherit the boundary problem unchanged.  That route is variational and
targets near-term hardware, whereas the explicit-circuit programme
pursued here targets fault-tolerant hardware with provable gate counts;
both need a forward model that does not reflect.  Classical
solvers handle this with absorbing layers, the de facto standard being the
perfectly matched layer (PML) and its convolutional,
complex-frequency-shifted variant
(CPML)~\cite{roden2000cpml,komatitsch2007unsplit,collino2001pml,duru2022pmlreview}.
Absorption is dissipation: the generator acquires a non-trivial Hermitian
part, and the dynamics ceases to be unitary.

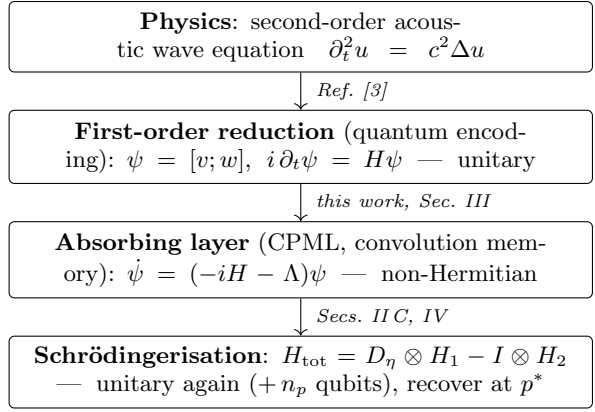
\begin{figure}[tb]
  \centering
  \begin{tikzpicture}[
      box/.style={draw, rounded corners=2pt, align=center,
                  inner sep=4pt, font=\small, text width=0.86\linewidth},
      lab/.style={font=\scriptsize\itshape, anchor=west},
      node distance=14pt]
    \node[box] (b1) {\textbf{Physics}: second-order acoustic wave
      equation\quad $\partial_t^2 u = c^2\Delta u$};
    \node[box, below=of b1] (b2) {\textbf{First-order reduction}
      (quantum encoding): $\psi=[v;w]$,\;
      $i\,\partial_t\psi = H\psi$ \;---\; unitary};
    \node[box, below=of b2] (b3) {\textbf{Absorbing layer} (CPML,
      convolution memory): $\dot\psi = (-iH-\Lambda)\psi$ \;---\;
      non-Hermitian};
    \node[box, below=of b3] (b4) {\textbf{Schr\"odingerisation}:
      $H_{\mathrm{tot}} = D_\eta\otimes H_1 - I\otimes H_2$ \;---\;
      unitary again ($+\,n_p$ qubits), recover at $p^{*}$};
    \draw[->] (b1) -- node[lab] {\;Ref.~\cite{sato2024hamiltonian}} (b2);
    \draw[->] (b2) -- node[lab] {\;this work, Sec.~\ref{sec:generators}} (b3);
    \draw[->] (b3) -- node[lab] {\;Secs.~\ref{sec:schro},
      \ref{sec:circuits}} (b4);
  \end{tikzpicture}
  \caption{The four-step structure: second-order acoustic physics,
  first-order quantum encoding, convolution absorbing layer applied in
  that first-order form, and Schr\"odingerisation restoring unitarity.
  The cost of step three---the indefiniteness quantified by
  Lemma~\ref{lem:threshold}---and its remedy (the Lyapunov symmetrizer)
  are the subject of Sec.~\ref{sec:symmetrizer}.}
  \label{fig:pipeline}
\end{figure}

Schr\"odingerisation~\cite{jin2023schrodingerisation} is a general route
for running such non-unitary linear dynamics on a quantum computer: the
warped phase-space transform turns $\dot z = Az$ into a genuine
Hamiltonian evolution on one extra register, at the price of a
post-selected recovery slice.  Figure~\ref{fig:pipeline} summarises the
resulting four-step structure of this work.

  The programme now spans artificial boundary
conditions for the time-dependent Schr\"odinger
equation~\cite{jin2024abc}, inhomogeneous terms and the recovery-threshold
theory~\cite{jin2024recovery}, ill-posed problems~\cite{jin2024illposed},
Maxwell's equations~\cite{jin2023maxwell,ma2024maxwellcircuits}, elastic
waves~\cite{jin2025elastic}, gate-level circuits for parabolic
problems~\cite{hu2024quantum,jin2024heat}, query-optimal
formulations~\cite{jin2025queries}, and implicit-explicit integrators
for multiscale dynamics including the bulk-damped telegraph
equation~\cite{hu2026imex}.  What it does not yet contain is an
absorbing boundary for a classical wave-type equation, at either matrix or
circuit level.

\subsection{Related work and positioning}
\label{sec:related}

The positioning below was verified by a full-text literature review
through June 2026.\footnote{Adversarially cross-checked sweep over
published and preprinted work, 2023--2026; all primary claims below were
confirmed against the full texts of the cited papers.}

\paragraph{The core niche is open.}  No published or preprinted work
through June 2026 combines any absorbing boundary (PML, CPML, complex
absorbing potential, sponge, or Dirichlet-to-Neumann map) for
\emph{classical wave-type equations} with Schr\"odingerisation or any
other unitarisation, at matrix or circuit level.  Jin and
Zhang~\cite{jin2025elastic} (elastic waves) state verbatim that
non-periodic boundaries ``remain an open issue'' left ``to the future.''
Maxwell circuits exist only for perfect-electric-conductor
boundaries~\cite{ma2024maxwellcircuits} and for impedance boundaries at
matrix level~\cite{jin2023maxwell}; the 2026 circuit
papers~\cite{liang2026structure,guseynov2026robin}
(Robin/Neumann/Dirichlet) contain no absorbing-boundary content.
Two 2025 works corroborate this independently.  B\"osch \emph{et
al.}~\cite{bosch2025sources}, simulating lossless wave equations by
Hamiltonian simulation, state that owing to their ``fundamentally
nonenergy conserving nature'' absorbing boundaries and PMLs are ``not
straightforward to implement\ldots within the HS framework,'' and resort
to approximate \emph{random} boundaries that scramble, rather than
absorb, outgoing energy.  Sato \emph{et
al.}~\cite{sato2025nonconservative} admit spatially varying bulk damping
$\zeta(x)\ge0$ in an LCHS formulation, but the resulting Hermitian part
is semidefinite by construction---the diagonal sponge/CAP case that
Lemma~\ref{lem:threshold} explicitly excludes---no layer-localised
damping or absorbing-layer construction appears, and no damped-wave
circuit is demonstrated.

\paragraph{PML plus Schr\"odingerisation per se is prior art} for the
time-dependent Schr\"odinger equation~\cite{jin2024abc}, at
query-complexity level and without circuits.  Our contribution is
therefore framed as \emph{wave-type equations plus explicit circuits}.

\paragraph{The recovery threshold is established theory}
\cite[Thm.~3.1]{jin2024recovery}, and an indefinite boundary-augmented
Hermitian part was already handled through that threshold for the heat
equation \cite[Remark~1]{jin2024heat}.  Our delta is the
\emph{structural} result: Lemma~\ref{lem:threshold} proves that the
indefiniteness of CPML memory forms cannot be removed by any diagonal
rescaling and quantifies its rate, $\Theta(\sqrt{\sigma_{\max}c/h})$.

\paragraph{Diagonal similarity preconditioning is prior art.}  Liang and
Liu~\cite{liang2026structure} use a diagonal transform to enforce a
positive-semidefinite Hermitian part before applying a
linear-combination-of-Hamiltonian-simulation algorithm to parabolic
problems.  This is the foil that sharpens our symmetrizer contribution:
Lemma~\ref{lem:threshold} proves that diagonal transforms are
\emph{insufficient} for CPML generators; the full (non-diagonal) Lyapunov
symmetrizer of Sec.~\ref{sec:symmetrizer} is both necessary and
sufficient.

\paragraph{Bell-basis circuits for non-unitary PDEs are prior art}
\cite{hu2024quantum,jin2024heat}.  Our circuit delta is narrower and is
framed accordingly: projector-valued operator strings for separated
multi-field (block) Hamiltonians, the two-level symmetrised splitting, and
the absorbing-boundary application at 17--21 qubits
(Sec.~\ref{sec:circuits}).

\subsection{Contributions}

Ordered by verified novelty:
\begin{enumerate}
\item the first absorbing boundaries for wave-type equations in any
  unitarisation framework, closing the open problem stated in
  Ref.~\cite{jin2025elastic};
\item the threshold lemma (Lemma~\ref{lem:threshold}) and the Lyapunov
  symmetrizer that circumvents it, with the diagonal-insufficiency
  contrast to Ref.~\cite{liang2026structure};
\item end-to-end validated explicit circuits, extending the Bell-basis
  constructions of Refs.~\cite{sato2024hamiltonian,hu2024quantum} to
  projector-valued block structure, with gold-standard classical
  validation.
\end{enumerate}

\section{Background}
\label{sec:background}

\subsection{The wave equation as Hamiltonian simulation}
\label{sec:wave}

Following Ref.~\cite{sato2024hamiltonian}, the wave equation
$\partial_t^2 u = c^2\Delta u$ is first-order-reduced through
$\psi = [v;\,w]$ with $v=\partial_t u$ and $w = i c\,\partial_x u$
(discretely, $w = icDu$), so that $i\,\partial_t\psi = H\psi$ with
\begin{equation}
  H \;=\; \begin{bmatrix} 0 & D \\ D^{\dagger} & 0 \end{bmatrix},
  \label{eq:waveH}
\end{equation}
where $D$ is the forward-difference operator and
$D^{\dagger}=-D_{\mathrm{b}}$ its negative backward counterpart.  Mixed
Dirichlet/Neumann boundary conditions emerge from the operator products
(the conventions of Ref.~\cite{sato2024hamiltonian}, verified numerically
in our implementation).  In two dimensions we use the separated-block form
$\psi = [v;\,w_x;\,w_y]$---one block per field, in the style of Eq.~(32)
of Ref.~\cite{sato2024hamiltonian} with $d=3$---instead of the
complex-combined two-block form, because per-direction damping requires
separated gradient components.

\subsection{The convolutional perfectly matched layer}
\label{sec:cpml}

The CPML stretches each coordinate by the complex-frequency-shifted (CFS)
factor
\begin{equation}
  s(\omega) \;=\; \kappa + \frac{\sigma}{\alpha + i\omega},
  \label{eq:cfs}
\end{equation}
which in the time domain replaces every spatial derivative by a causal
convolution with an exponential kernel~\cite{roden2000cpml},
\begin{equation}
  \partial_x f \;\longrightarrow\; \frac{1}{\kappa}\,\partial_x f
    + \zeta * \partial_x f, \qquad
  \zeta(t) = -\frac{\sigma}{\kappa^{2}}\,
    e^{-(\sigma/\kappa+\alpha)t}\,\theta(t).
  \label{eq:kernel}
\end{equation}
Each convolution is carried by one memory field obeying the local ODE
\begin{equation}
  \partial_t \phi \;=\; -\Bigl(\frac{\sigma}{\kappa}+\alpha\Bigr)\phi
    \;-\; \frac{\sigma}{\kappa^{2}}\,\partial_x f .
  \label{eq:memoryode}
\end{equation}
The damping profile is polynomially graded inside a layer of
$n_{\mathrm{pml}}$ points, $\sigma(d) = \sigma_{\max}(d/L)^{m}$ with
$\sigma_{\max} = -(m{+}1)\,c\ln R_0/(2L)$~\cite{collino2001pml}; we use
the standard recipe $m=2$, $R_0=10^{-3}$ with the parameter conventions of
Komatitsch and Martin~\cite{komatitsch2007unsplit}.  The profile is
sampled at the staggered locations where each field effectively lives
($v$ at integer nodes, $w$ at half-cells).  This stagger is load-bearing:
an ablation shows that unstaggered sampling degrades the measured
reflection by factors of 120--235.

\subsection{Schr\"odingerisation}
\label{sec:schro}

Write the semi-discrete absorbing dynamics as $\dot z = Az$ \footnote{This is the a linear ODE, need to find z(t) given initial condition z(0); z is Psi in this case} and split
$A = H_1 + iH_2$ with $H_1=(A+A^{\dagger})/2$ and
$H_2=(A-A^{\dagger})/(2i)$, both Hermitian \footnote{Correct for all squared A}.  The warped transform
$w(t,p) = e^{-p}z(t)$ (for $p\ge 0$, extended by a profile $g$ below)
obeys $\partial_t w = -H_1\partial_p w + iH_2 w$, and a Fourier series in
$p$ decouples it into unitary evolutions per mode
\cite{jin2023schrodingerisation},
\begin{equation}
  \hat w_k(T) \;=\; e^{-iT(\eta_k H_1 - H_2)}\,\hat w_k(0),
  \label{eq:permode}
\end{equation}
i.e.\ a Hamiltonian simulation of
$H_{\mathrm{tot}} = D_\eta\otimes H_1 - I\otimes H_2$ on the system
register plus one $p$ register of $n_p$ qubits.  The initial data
$w(0,p)=g(p)\,z(0)$ is a product state across the two registers.  The
solution is recovered from a slice,
\begin{equation}
  z(T) \;=\; e^{p^{*}} w(T,p^{*}),
  \qquad
  p^{*} \;\ge\; \lambda^{+}_{\max}(H_1)\,T,
  \label{eq:recovery}
\end{equation}
where the threshold on $p^{*}$ is Theorem~3.1 of
Ref.~\cite{jin2024recovery}.  The kinked profile $g(p)=e^{-|p|}$ is only
$O(\Delta p)$ accurate; the $C^{1}$ cubic profile of
\cite[Remark~4.9]{jin2024recovery} restores $O(\Delta p^{2})$ while
keeping $g(p)=e^{-p}$ for $p\ge0$, leaving Eq.~\eqref{eq:recovery}
unchanged.

\section{Absorbing generators}
\label{sec:generators}

\subsection{One dimension: exact collapse of the convolution}
\label{sec:1d}

With $\kappa=1$, $\alpha=0$ the convolution collapses \emph{exactly} (for
data supported outside the layer) to block-diagonal damping \footnote{Sato: $A=-iH$, closed system},
\begin{equation}
  \dot\psi = A\psi, \qquad
  A \;=\; -iH - \diag(\sigma_v,\, \sigma_w),
  \label{eq:1dcollapse}
\end{equation}
so $H_1 = -\diag(\sigma)\preceq 0$ and the recovery
threshold~\eqref{eq:recovery} is unconditional ($\lambda^{+}=0$, any small
$p^{*}>0$ is valid).  The proof is the memory ansatz
$\phi = -\sigma v/\gamma$, which satisfies the memory
ODE~\eqref{eq:memoryode} identically for data supported outside the
layer; we verify the collapse numerically to $6\times10^{-13}$.  The
general CFS memory form (two extra block qubits) is provided for
$\kappa>1$ or $\alpha>0$.

\subsection{Two dimensions: memory form and classical validation}
\label{sec:2d}

In 2D the damping acts per direction, so the state uses the
separated-block form padded to eight blocks (three extra block qubits),
\begin{equation}
  \bigl[\,v;\; w_x;\; w_y;\; \phi_{vx};\; \phi_{vy};\;
        \phi_{x};\; \phi_{y};\; 0\,\bigr],
  \label{eq:blocks}
\end{equation}
with the memory injections masked to the layer strips.  Corners need no
special treatment---both directions' memory fields are simultaneously
active, the structural advantage of unsplit CPML over split
formulations~\cite{roden2000cpml}.  Long-time stability is a serious
concern for PMLs in
general~\cite{becache2003stability,becache2004longtime,duru2022pmlreview};
here we measure $\max\operatorname{Re}\operatorname{eig}(A) = 0$ to
round-off across all tested configurations.

\emph{Gold-standard validation.}  Comparing the truncated domain against
a $4\times$ enlarged reference, the true reflection is $3.4\times10^{-4}$
(1D, $n_{\mathrm{pml}}=12$) and $2.8\times10^{-4}$ (2D $32\times32$,
$n_{\mathrm{pml}}=8$), at the floor reported by Komatitsch and
Martin~\cite{komatitsch2007unsplit}.

\begin{figure*}
  \includegraphics[width=\textwidth]{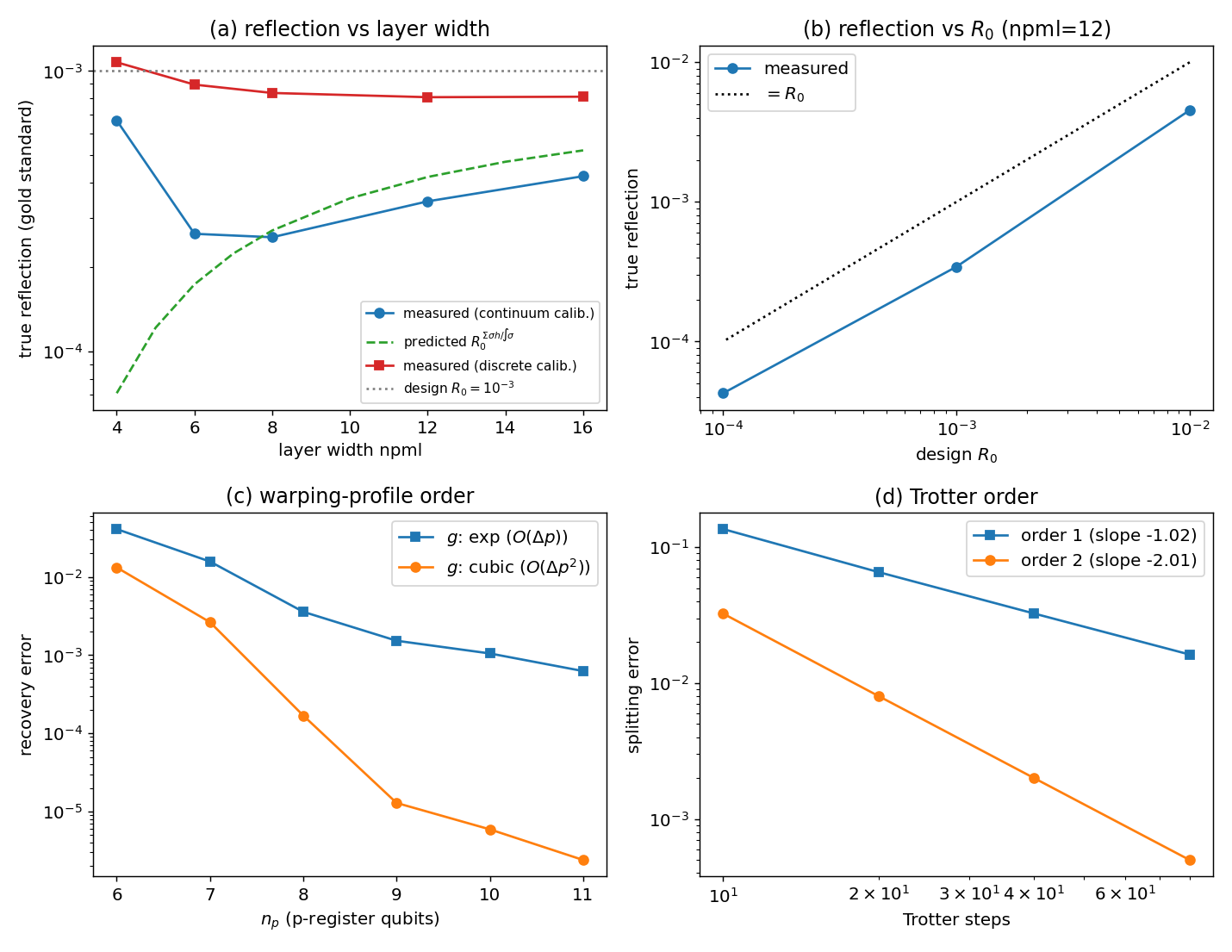}
  \caption{Classical and Schr\"odingerisation convergence.
  (a)~Gold-standard reflection (truncated domain versus $4\times$
  reference) versus layer width $n_{\mathrm{pml}}$ at design reflection
  $R_0=10^{-3}$.  The curve is non-monotonic: the transition error falls
  with layer width while the wall return rises toward the design $R_0$
  from below, because node sampling of the polynomial profile
  over-damps thin layers (dashed curve: the predicted discrete
  wall-return $R_{\mathrm{eff}}=R_0^{\,\Sigma\sigma
  h/\!\int\!\sigma}$, which the measurement tracks within
  $5$--$20\%$ for $n_{\mathrm{pml}}\ge8$).  Red squares: the
  discretely calibrated profile pair (Appendix~\ref{appD:validation}),
  for which the realised reflection is flat at the design value.
  (b)~Reflection versus $R_0$ at $n_{\mathrm{pml}}=12$: the measured
  reflection tracks the Collino--Tsogka design value within a factor
  ${\sim}2.3$ over two decades.
  (c)~Recovery error versus $p$-register size $n_p$ for the kinked
  $e^{-|p|}$ warping profile [$O(\Delta p)$] and the $C^{1}$ cubic
  profile [$O(\Delta p^{2})$] of Ref.~\cite{jin2024recovery}, 1D CPML
  benchmark.
  (d)~Trotter splitting error versus step count at orders 1 and 2
  (fitted slopes $-1.02$ and $-2.01$), measured against the exact
  Schr\"odingerised evolution to isolate the splitting error.}
  \label{fig:convergence}
\end{figure*}

\emph{Convergence} [1D gold standard, $N=128$ versus an $N=512$
reference, plateau over $T\in\{100,120,140\}$ with plateau spread
$<10^{-9}$, i.e.\ fully converged; Figs.~\ref{fig:convergence}(a,b)]:
\begin{itemize}
\item versus $n_{\mathrm{pml}}$ at $R_0=10^{-3}$: $6.7\times10^{-4}$ (4),
  $2.6\times10^{-4}$ (6), $2.6\times10^{-4}$ (8), $3.4\times10^{-4}$ (12),
  $4.2\times10^{-4}$ (16)---\emph{non-monotonic}, with a minimum at
  $n_{\mathrm{pml}}\approx6$--$8$.  The two contributions are the
  discrete transition error, which decreases with layer width, and the
  wall-return reflection, which \emph{increases toward the design value
  from below}: node sampling of the polynomial profile overestimates
  $\int\sigma\,dx$ by a factor $1+O(1/n_{\mathrm{pml}})$, so the
  effective round-trip reflection is
  $R_{\mathrm{eff}}=R_0^{\,\Sigma\sigma h/\!\int\!\sigma}$
  ($R_0^{1.38}=7.1\times10^{-5}$ at $n_{\mathrm{pml}}=4$ rising to
  $R_0^{1.09}=5.2\times10^{-4}$ at $16$); for
  $n_{\mathrm{pml}}\ge8$ the measured reflection equals this
  prediction within $5$--$20\%$, i.e.\ the layer performs at design,
  converging to $R_0$ from below [Fig.~\ref{fig:convergence}(a)].
  Jointly rescaling the staggered profile pair so the discrete sum
  matches the design integral (\texttt{calibration='discrete'},
  Appendix~\ref{appD:validation}) makes the realised reflection flat at
  $0.81$--$1.08\,R_0$ over the whole sweep---a one-line calibration fix
  that applies to classical CPML codes as well;
\item versus $R_0$ at $n_{\mathrm{pml}}=12$: $4.5\times10^{-3}$
  ($10^{-2}$), $3.4\times10^{-4}$ ($10^{-3}$), $4.3\times10^{-5}$
  ($10^{-4}$)---tracking the design reflection $R_0$ within a factor
  ${\sim}2.3$ over two decades.  The Collino--Tsogka amplitude
  formula~\cite{collino2001pml} is predictive in this discretisation.
\end{itemize}

\subsection{The recovery-threshold obstruction}
\label{sec:lemma}

The memory form pays a price.  The injection column ($\gamma$ on one
side) and the memory-source row ($\sigma\partial_x/\gamma$ on the other)
of each memory field are structurally asymmetric, and this asymmetry
cannot be scaled away.  Throughout this section
$\beta_j = \sigma_j/\kappa_j+\alpha_j\ge0$ denotes the CFS decay rate at
node $j$; we set the wave speed $c=1$ (so the stencil entries obey
$|D_{ji}|=1/h$), the general rate restoring a factor of $c$.

\begin{lemma}[diagonal irreducibility]
\label{lem:threshold}
Let $A$ be the memory-form CPML generator of Sec.~\ref{sec:2d} and let
$G=\diag(\gamma_j)>0$ be any diagonal rescaling of the memory fields.
Let $s_j$ denote the magnitude of the memory-source coupling
($s_j\sim\sigma_j/(\gamma_j h)$ for the difference stencil).  Then the
Hermitian part $H_1(G)$ of the rescaled generator satisfies
\begin{equation}
  \lamax\bigl(H_1(G)\bigr) \;\ge\;
  \max_j\, \max\bigl\{\, f(\gamma_j,\beta_j),\; g(s_j,\beta_j) \,\bigr\}
  \;>\; 0,
  \label{eq:lemmabound}
\end{equation}
with $f(\gamma,\beta) = \bigl(\sqrt{\beta^{2}+\gamma^{2}}-\beta\bigr)/2$,
arising from the $2\times2$ principal submatrix
$\bigl[\begin{smallmatrix} 0 & \gamma/2 \\ \gamma/2 & -\beta
\end{smallmatrix}\bigr]$
on a $(v_j,\phi_j)$ pair by Cauchy interlacing, and
$g(s,\beta) = f(s,\beta)$ from a $(w_i,\phi_j)$ pair (the one-directional
coupling entry $s$ enters $H_1$ as $s/2$, so both pairs are governed by
the same $f$).  Optimising $\gamma$ against the two bounds gives
\begin{equation}
  \lamax(H_1) \;=\; \Theta\!\left(\sqrt{\sigma_{\max}\,c/h}\right)
  \qquad \text{as } h\to0,
  \label{eq:lemmarate}
\end{equation}
independent of the scaling.
\end{lemma}

The full proof is given in Appendix~\ref{app:proof}.  Its consistency
check is instructive: the collapsed 1D form of Sec.~\ref{sec:1d} has
\emph{no} injection pair (the memory is eliminated and the damping is
diagonal), so the lemma does not apply---and indeed
$H_1=-\diag(\sigma)\preceq0$ there.

\begin{figure}
  \includegraphics[width=\columnwidth]{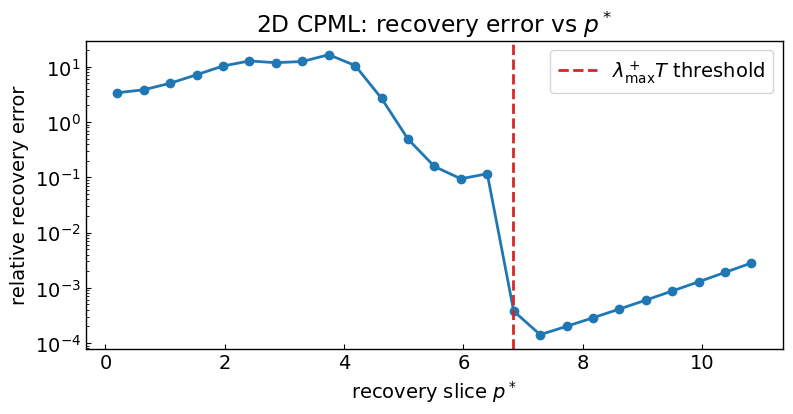}
  \caption{The recovery threshold in action (2D CPML, $8\times8$ grid,
  $\sigma_{\max}=0.5$, $T=10$).  Relative recovery error versus the slice
  position $p^{*}$: below the threshold $p^{*}=\lambda^{+}T$ (dashed
  line, $\lambda^{+}=0.683$) the recovered state is $O(1)$ wrong; just
  above it the error drops by four orders of magnitude, then grows
  slowly as $e^{p^{*}}$ amplifies the $p$-grid discretisation floor.}
  \label{fig:threshold}
\end{figure}

\emph{Consequence.}  By the threshold of Eq.~\eqref{eq:recovery}
\cite[Thm.~3.1]{jin2024recovery}, recovery requires
$p^{*}\ge\lambda^{+}T$ with $\lambda^{+}\equiv\lamax^{+}(H_1)$, and the
post-selection probability of the recovery window scales as
$e^{-2p^{*}} = e^{-2\lambda^{+}T}$.  Numerically: at $\sigma_{\max}=0.5$
on an $8\times8$ grid, $\lambda^{+}=0.683$; recovery below the threshold
is $O(1)$ wrong, while above it the error drops to $2\times10^{-3}$
relative ($10^{-4}$ with $n_p=11$); see Fig.~\ref{fig:threshold}.  The
sponge/CAP alternative (diagonal damping $\sigma_x+\sigma_y$, four
blocks) has $H_1\preceq0$ unconditionally at the price of matched-layer
quality.

\input{sections/cpml_vs_sponge}

\subsection{The Lyapunov symmetrizer}
\label{sec:symmetrizer}

A matrix $W>0$ solving the Lyapunov inequality
$WA + A^{\dagger}W\preceq0$ gives, with $S=W^{1/2}$, a transformed
generator $\At = SAS^{-1}$ whose Hermitian part satisfies
\begin{equation}
  \Ht \;=\; \tfrac12\, S^{-1}\bigl(WA + A^{\dagger}W\bigr)S^{-1}
  \;\preceq\; 0
  \label{eq:congruence}
\end{equation}
by congruence.  Recovery of $\tilde z = Sz$ is then valid at any
$p^{*}>0$, and $z = S^{-1}\tilde z$ is recovered classically (or folded
into the readout).  Since $A$ is only marginally stable (interior modes
are neutral), we construct $W$ from the $\epsilon$-shifted generator:
\begin{equation}
  A_\epsilon = A - \epsilon I, \qquad
  A_\epsilon^{\dagger}W + WA_\epsilon = -I, \qquad
  S = W^{1/2},
  \label{eq:lyapunov}
\end{equation}
the shift being undone classically by the factor $e^{\epsilon T}$.  An
exact identity follows (Proposition~\ref{appE:prop} in
Appendix~\ref{appE:symmetrizer}: congruence gives
$\mathrm{Herm}(SA_\epsilon S^{-1})=-W^{-1}/2$ exactly; verified to better
than $10^{-11}$ relative in all 18 measured configurations): the
transform of the \emph{original} $A$ has
\begin{equation}
  \lamax(\Ht) \;=\; \epsilon - \frac{1}{2\lamax(W)} \;\approx\; \epsilon,
  \label{eq:identity}
\end{equation}
while the transform of $A_\epsilon$ has $\Ht\preceq0$ strictly.  We
therefore use the shifted transform; the undo factor
$e^{\epsilon T}\le e^{0.4}$ over the horizons tested is trivial.

\begin{table*}
\caption{Lyapunov-symmetrizer conditioning across all measured
configurations [grids $(n_x,n_y)$ qubits per dimension, i.e.\ $8\times8$,
$16\times8$, $16\times16$ points; $n$ is the full state dimension
including the eight blocks].  $\lambda^{+}=\lamax^{+}(H_1)$ is the
indefiniteness of the untransformed generator;
$\kappa_2(S)=\sqrt{\operatorname{cond}_2(W)}$ the conditioning of the
symmetrizer; $\lamax(\Ht^{(\epsilon)})$ the (strictly negative) largest
Hermitian-part eigenvalue of the $\epsilon$-shifted transform
$SA_\epsilon S^{-1}$; and $T^{*}=\ln\kappa^{2}/(2\lambda^{+})$ the
crossover time beyond which the symmetrizer beats the direct
above-threshold recovery.  All Lyapunov residuals are
$\le1.4\times10^{-14}$ (relative).}
\label{tab:kappa}
\begin{ruledtabular}
\begin{tabular}{llllllll}
Grid & $n$ & $\sigma_{\max}$ & $\epsilon$ & $\lambda^{+}(H_1)$ &
$\kappa_2(S)$ & $\lamax(\Ht^{(\epsilon)})$ & $T^{*}$ \\
\colrule
$8\times8$ & 512 & 0.5 & $10^{-2}$ & 0.683 & $4.0\times10^{2}$ & $-4.6\times10^{-6}$ & 8.8 \\
$8\times8$ & 512 & 0.5 & $10^{-3}$ & 0.683 & $3.0\times10^{3}$ & $-8.1\times10^{-8}$ & 11.7 \\
$8\times8$ & 512 & 0.5 & $10^{-4}$ & 0.683 & $1.1\times10^{4}$ & $-6.0\times10^{-9}$ & 13.6 \\
$8\times8$ & 512 & 1.0 & $10^{-2}$ & 0.927 & $5.1\times10^{2}$ & $-5.2\times10^{-6}$ & 6.7 \\
$8\times8$ & 512 & 1.0 & $10^{-3}$ & 0.927 & $2.9\times10^{3}$ & $-1.6\times10^{-7}$ & 8.6 \\
$8\times8$ & 512 & 1.0 & $10^{-4}$ & 0.927 & $9.8\times10^{3}$ & $-1.4\times10^{-8}$ & 9.9 \\
$16\times8$ & 1024 & 0.5 & $10^{-2}$ & 0.687 & $5.1\times10^{2}$ & $-3.0\times10^{-6}$ & 9.1 \\
$16\times8$ & 1024 & 0.5 & $10^{-3}$ & 0.687 & $7.0\times10^{3}$ & $-1.5\times10^{-8}$ & 12.9 \\
$16\times8$ & 1024 & 0.5 & $10^{-4}$ & 0.687 & $3.1\times10^{4}$ & $-8.0\times10^{-10}$ & 15.0 \\
$16\times8$ & 1024 & 1.0 & $10^{-2}$ & 0.956 & $7.5\times10^{2}$ & $-2.4\times10^{-6}$ & 6.9 \\
$16\times8$ & 1024 & 1.0 & $10^{-3}$ & 0.956 & $7.0\times10^{3}$ & $-2.8\times10^{-8}$ & 9.3 \\
$16\times8$ & 1024 & 1.0 & $10^{-4}$ & 0.956 & $2.7\times10^{4}$ & $-1.8\times10^{-9}$ & 10.7 \\
$16\times16$ & 2048 & 0.5 & $10^{-2}$ & 0.694 & $5.7\times10^{2}$ & $-2.4\times10^{-6}$ & 9.1 \\
$16\times16$ & 2048 & 0.5 & $10^{-3}$ & 0.694 & $9.9\times10^{3}$ & $-7.7\times10^{-9}$ & 13.3 \\
$16\times16$ & 2048 & 0.5 & $10^{-4}$ & 0.694 & $6.6\times10^{4}$ & $-1.7\times10^{-10}$ & 16.0 \\
$16\times16$ & 2048 & 1.0 & $10^{-2}$ & 0.972 & $8.6\times10^{2}$ & $-1.9\times10^{-6}$ & 6.9 \\
$16\times16$ & 2048 & 1.0 & $10^{-3}$ & 0.972 & $1.2\times10^{4}$ & $-8.7\times10^{-9}$ & 9.7 \\
$16\times16$ & 2048 & 1.0 & $10^{-4}$ & 0.972 & $6.3\times10^{4}$ & $-3.4\times10^{-10}$ & 11.4 \\
\end{tabular}
\end{ruledtabular}
\end{table*}

\begin{table*}
\caption{End-to-end recovery tradeoff: direct versus symmetrized
recovery, errors
$\lVert z_{\mathrm{rec}}-e^{AT}z_0\rVert_2/\lVert e^{AT}z_0\rVert_2$.
``Small $p^{*}$'' is the library-default slice three grid points above
zero.  For the untransformed system above threshold,
$p^{*}=\lambda^{+}T+1$ and the post-selection penalty is $e^{2p^{*}}$;
entries marked ``--'' exceed the $p$-domain
($\lambda^{+}T+1>p_{\max}$).  The symmetrized column uses the
$\epsilon$-shifted transform at $\epsilon=10^{-3}$ recovered at the same
small $p^{*}$, with the $T$-independent penalty
$\kappa_2^{2}(S)$ of Table~\ref{tab:kappa}.  The $16\times16$ run uses a
reduced $p$ register ($n_p=8$), which sets its higher error floor.}
\label{tab:tradeoff}
\begin{ruledtabular}
\begin{tabular}{llllllll}
 & & \multicolumn{4}{c}{Untransformed} & \multicolumn{1}{c}{Symmetrized ($\epsilon=10^{-3}$)} \\
\cline{3-6}\cline{7-7}
Configuration & $T$ & err.\ (small $p^{*}$) & $p^{*}\!\ge\!\lambda^{+}T$ &
err.\ (above thr.) & penalty $e^{2p^{*}}$ & err.\ (small $p^{*}$) \\
\colrule
$8\times8$ ($\sigma_{\max}=0.5$, $n_p=11$) & 10 & $3.4$ & 7.8 & $1.2\times10^{-4}$ & $6.3\times10^{6}$ & $1.2\times10^{-6}$ \\
 & 20 & $8.3$ & 14.7 & $2.4\times10^{-1}$ & $5.4\times10^{12}$ & $2.4\times10^{-6}$ \\
 & 40 & $2.6\times10^{1}$ & -- & -- & -- & $4.4\times10^{-6}$ \\
$8\times8$ ($\sigma_{\max}=1.0$, $n_p=11$) & 10 & $7.2$ & 10.3 & $2.2\times10^{-2}$ & $8.4\times10^{8}$ & $2.1\times10^{-5}$ \\
 & 20 & $1.6\times10^{1}$ & 19.6 & $3.0\times10^{2}$ & $1.0\times10^{17}$ & $2.3\times10^{-5}$ \\
 & 40 & $2.9\times10^{1}$ & 38.1 & $7.3\times10^{15}$ & $1.2\times10^{33}$ & $3.5\times10^{-5}$ \\
$16\times8$ ($\sigma_{\max}=0.5$, $n_p=10$) & 10 & $2.5$ & 7.9 & $1.1\times10^{-4}$ & $7.4\times10^{6}$ & $4.2\times10^{-6}$ \\
 & 20 & $4.8$ & 14.8 & $4.5\times10^{-1}$ & $6.7\times10^{12}$ & $6.0\times10^{-6}$ \\
 & 40 & $1.3\times10^{1}$ & -- & -- & -- & $1.1\times10^{-5}$ \\
$16\times16$ ($\sigma_{\max}=0.5$, $n_p=8$) & 10 & $2.0$ & 8.0 & $1.3\times10^{-2}$ & $9.2\times10^{6}$ & $1.1\times10^{-4}$ \\
 & 20 & $5.6$ & 15.0 & $7.5\times10^{1}$ & $1.0\times10^{13}$ & $5.2\times10^{-4}$ \\
 & 40 & $2.5\times10^{1}$ & -- & -- & -- & $3.8\times10^{-3}$ \\
\end{tabular}
\end{ruledtabular}
\end{table*}

\begin{figure*}
  \includegraphics[width=\textwidth]{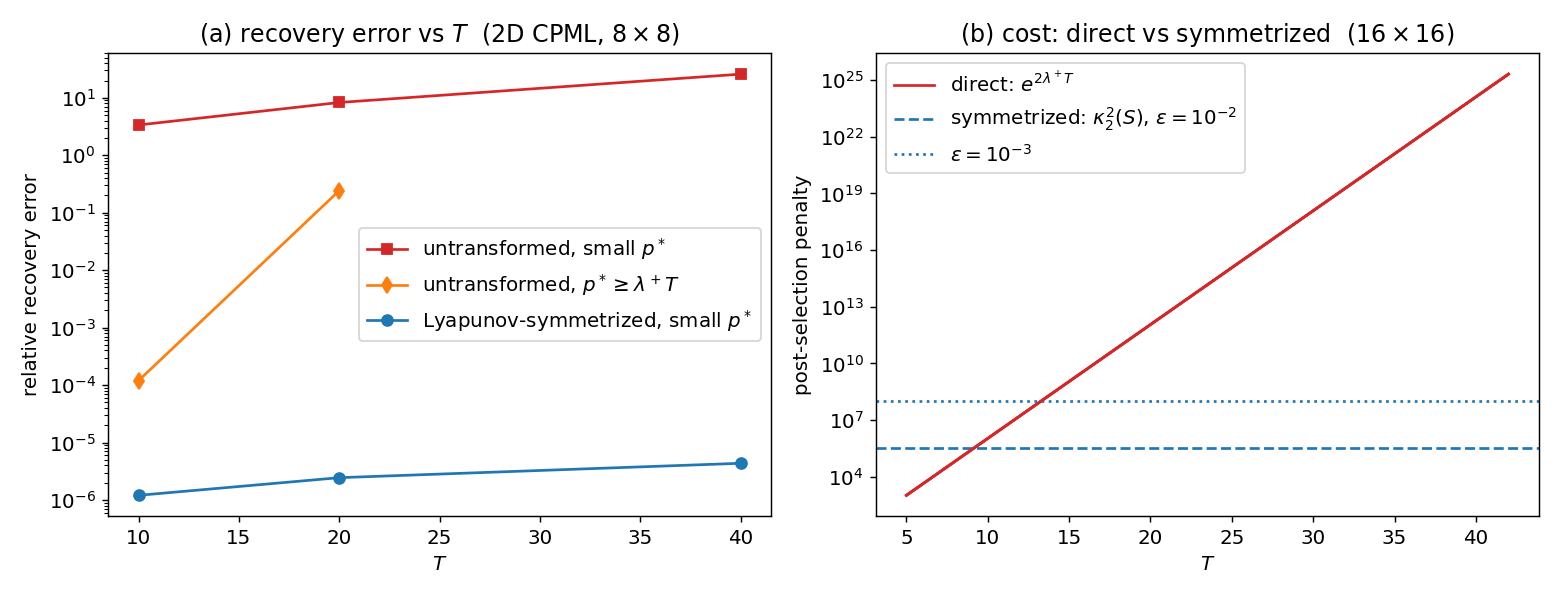}
  \caption{The Lyapunov symmetrizer.  (a)~Relative recovery error versus
  evolution time $T$ for the 2D CPML system ($8\times8$ grid,
  $\sigma_{\max}=0.5$): untransformed recovery at the default small
  $p^{*}$ (squares) is $O(1)$ garbage; untransformed recovery above the
  threshold $p^{*}\ge\lambda^{+}T$ (diamonds) degrades with $T$ as
  $e^{p^{*}}$ amplifies the $p$-grid floor and leaves the admissible
  $p$-domain entirely by $T=40$; the Lyapunov-symmetrized recovery at the
  same small $p^{*}$ (circles, $\epsilon=10^{-3}$, shifted transform)
  stays at the $10^{-6}$ level, essentially flat in $T$.
  (b)~Post-selection cost on the $16\times16$ grid: the direct
  $e^{2\lambda^{+}T}$ penalty (solid) against the $T$-independent
  symmetrizer penalty $\kappa_2^{2}(S)$ at $\epsilon=10^{-2}$ (dashed)
  and $10^{-3}$ (dotted); the crossovers are the $T^{*}$ values of
  Table~\ref{tab:kappa}.}
  \label{fig:symmetrizer}
\end{figure*}

\emph{Measured results} [grids $8\times8$, $16\times8$, $16\times16$
($n=512$--$2048$), $\sigma_{\max}\in\{0.5,1.0\}$,
$\epsilon\in\{10^{-2},10^{-3},10^{-4}\}$; Lyapunov residuals
$\le1.4\times10^{-14}$; Tables~\ref{tab:kappa} and \ref{tab:tradeoff},
Fig.~\ref{fig:symmetrizer}]:
\begin{itemize}
\item \emph{End-to-end recovery at the default small $p^{*}$}
  ($\epsilon=10^{-3}$): transformed errors
  $1.2\times10^{-6}$--$3.5\times10^{-5}$ on the configurations with
  $n_p\in\{10,11\}$, essentially flat in $T$ over $T\in\{10,20,40\}$; the
  $16\times16$ run with a reduced $p$ register ($n_p=8$) gives
  $1.1\times10^{-4}$--$3.8\times10^{-3}$, consistent with its
  $O(\Delta p^{2})$ floor.  The untransformed system at the same $p^{*}$
  returns $O(1)$--$O(10)$ garbage growing with $T$.
\item \emph{The direct above-threshold method is worse than its nominal
  $e^{2\lambda^{+}T}$ cost}: at $T=20$ it degrades (errors $0.24$--$0.45$
  at $\sigma_{\max}=0.5$) and at $T=40$ it explodes (error up to
  $7\times10^{15}$ at $p^{*}=38$) because $e^{p^{*}}$ amplifies the
  $p$-grid floor.  At long $T$ the symmetrizer is not merely cheaper---it
  is the only working recovery.
\item \emph{Conditioning}:
  $\kappa_2(S)=4.0\times10^{2}$--$8.6\times10^{2}$ at $\epsilon=10^{-2}$
  (the recommended setting), flat in $\sigma_{\max}$, scaling as
  ${\sim}\epsilon^{-0.6\,\text{to}\,-1.0}$---and \emph{saturating} in
  $n$: extending the measurement to $n=4096$ gives
  $\kappa_2(S)=564.5$ at $\epsilon=10^{-2}$ versus $566.9$ at $n=2048$
  (local slope $-0.006$; over the full range $n=512$--$4096$ the fitted
  power-law exponent drops from $0.25$ to $0.16$ with degrading fit
  quality, i.e.\ a power law is the wrong model because $\kappa$ is
  flattening).  $\lambda^{+}$ itself saturates ($0.683\to0.694$ over
  $8\times$ in $n$) and $\lambda_{\min}(W)$ stays at $0.66$--$0.68$, so
  the $\kappa$ growth is driven entirely by
  $\lambda_{\max}(W)\sim1/(2\epsilon)$: the conditioning is
  $\epsilon$-controlled, not grid-controlled, and the symmetrizer cost
  does not blow up under mesh refinement.  The eigenvector symmetrizer
  $W=(VV^{\dagger})^{-1}$ is four to six orders of magnitude worse
  conditioned ($\kappa_2\sim3\times10^{8}$) and fails outright on
  near-defective configurations---the Lyapunov route is the right
  construction.
\item \emph{Tradeoff}:
  $T^{*}=\ln\kappa^{2}/(2\lambda^{+})=6.7$--$16.0$ across all
  configurations; at $T=20$ the symmetrizer wins by 4--9 orders of
  magnitude in post-selection cost, at $T=40$ by 16--26 orders.  $T^{*}$
  grows only logarithmically with $\kappa$, hence very weakly with grid
  size.
\end{itemize}

\emph{Structure and compressibility of $S$.}  We measured the structure
of $S$ directly and tested 25 approximate symmetrizers (distance-banded,
magnitude-thresholded, block-pattern, identity-plus-low-rank, hybrid,
and inverse-side-banded families) for dissipativity and end-to-end
recovery on the $n=512$ and $n=1024$ systems.  Three findings.
(i)~Identity-plus-low-rank is \emph{ruled out}: the correction
$E=S-\alpha I$ is numerically full rank (995 of 1024 singular values
above $10^{-2}$ of the top), and $\lamax(\tilde H_1)$ stays at the
\emph{untransformed} $\lambda^{+}=0.67$ for ranks $8$--$64$---no
few-term LCU implementation of $S$ exists.  (ii)~Strict sparsification
preserving $\lamax(\tilde H_1)\le\epsilon$ requires essentially the
full active support ($\approx48$--$51\%$ of entries; on the remaining
``dead'' indices---interior memory and padding, $30$--$50\%$ of the
space---$S$ is exactly a multiple of the identity, which is free
structure).  (iii)~\emph{Moderate compression works} once the residual
$\lambda_b$ of an approximate $S_b$ is absorbed as an additional shift
(undone by $e^{\lambda_b T}\le e^{2.6}\approx14$ at the worst measured
$\lambda_b=0.066$, $T=40$): magnitude thresholding at $35\%$ density
recovers to $8.7\times10^{-6}$ ($T=10$) and $5.4\times10^{-5}$
($T=40$); a rank-$32$-plus-band-$8$ hybrid at $26\%$ reaches
$1.3\times10^{-5}$ and $6.6\times10^{-4}$ (all $n=1024$; the
conditioning of every usable $S_b$ stays ${\sim}5\times10^{2}$, equal
to the exact $S$).  The energy of $E$ is $99.6\%$ concentrated on
layer-strip index pairs.  The implementation route is therefore
block-encoding of a ${\sim}35\%$-dense operator (times the free
dead-index identity factor), not a local circuit or a short LCU; at
current theory the symmetrizer is best used as classical pre/\allowbreak
post-processing, and a structured-ansatz Lyapunov construction (solving
for $W$ within a polynomially parametrised operator family) is the open
problem these measurements sharpen.

\emph{Honest limitations.}  The construction requires a dense $O(n^{3})$
classical precomputation of $W$, $S$, $S^{-1}$; the fate of the sparse
quantised-tensor-train (QTT) term structure under the dense similarity
(Trotter circuits for $\At$) is not assessed; the $\kappa$ scaling
beyond $n=4096$ is extrapolation (though saturating); the compression
study is at $n\le1024$.

\section{Explicit circuits}
\label{sec:circuits}

\subsection{Projector-extended Bell-basis term evolution}
\label{sec:projector}

\begin{figure}[tb]
  \centering
  \makebox[0pt][r]{(a)\hspace{0.96\linewidth}}\\[-0.5ex]
  \resizebox{\linewidth}{!}{%
  \begin{quantikz}[row sep={0.62cm,between origins}, column sep=0.22cm]
    \lstick{$q_0\;(\sigma_{10})$} & \qw & \targ{} & \qw & \qw
      & \ctrl{1} & \qw & \qw & \targ{} & \qw & \qw \\
    \lstick{$q_1\;(\sigma_{01})$} & \targ{} & \qw & \qw & \qw
      & \octrl{3} & \qw & \qw & \qw & \targ{} & \qw \\
    \lstick{$q_2\;(I)$} & \qw & \qw & \qw & \qw
      & \qw & \qw & \qw & \qw & \qw & \qw \\
    \lstick{$q_3\;(I)$} & \qw & \qw & \qw & \qw
      & \qw & \qw & \qw & \qw & \qw & \qw \\
    \lstick{$q_4\;(\sigma_{00})$} & \qw & \qw & \qw & \qw
      & \octrl{1} & \qw & \qw & \qw & \qw & \qw \\
    \lstick{$q_5\;(\sigma_{01},\,\mathrm{piv.})$}
      & \ctrl{-4} & \ctrl{-5} & \gate{P(\lambda)} & \gate{H}
      & \gate{R_z(2\gamma\,\delta t)} & \gate{H} & \gate{P(-\lambda)}
      & \ctrl{-5} & \ctrl{-4} & \qw
  \end{quantikz}}\\[1.2ex]
  \makebox[0pt][r]{(b)\hspace{0.96\linewidth}}\\[-0.5ex]
  \resizebox{\linewidth}{!}{%
  \begin{quantikz}[row sep={0.95cm,between origins}, column sep=0.18cm]
    \lstick{sys} & \qwbundle{n_s} & \gate{\lvert\psi_0\rangle} & \qw
      & \gate[2]{e^{+i\frac{\Delta t}{2}\eta\Sigma}}
      & \gate{e^{-i\Delta t H}}
      & \gate[2]{e^{+i\Delta t\,\eta\Sigma}}
      & \ \ldots\
      & \gate{e^{-i\Delta t H}}
      & \gate[2]{e^{+i\frac{\Delta t}{2}\eta\Sigma}}
      & \qw & \qw \\
    \lstick{$p$} & \qwbundle{n_p} & \gate{\lvert g\rangle}
      & \gate{\mathrm{QFT}^{\dagger}} & {} & \qw & {}
      & \ \ldots\
      & \qw & {} & \gate{\mathrm{QFT}} & \meter{p^{*}}
  \end{quantikz}}
  \caption{Explicit circuits.  (a)~Bell-basis evolution
  $e^{-i\,\delta t\,(cP+\mathrm{h.c.})}$ of the projector-containing
  term $P=\sigma_{01}\otimes\sigma_{00}\otimes I\otimes I\otimes
  \sigma_{01}\otimes\sigma_{10}$ of the separated 2D wave Hamiltonian
  (string \texttt{mzIImp}, qubit $q_k$ carrying string character $5-k$;
  $\lambda=\arg c$, $\gamma=|c|$; $\lambda=0$ for the real wave-equation
  coefficients): \textsc{cnot}s rotate the ladder qubits to the Bell
  basis, and the single multi-controlled $R_z$ carries the block
  projector $\sigma_{00}$ ($q_4$) as an \emph{open-circle}
  $0$-control---the entire compilation rule of
  Sec.~\ref{sec:projector}.  The gate sequence is unitarily identical to
  the library-compiled circuit (verified to machine precision).
  (b)~The Schr\"odingerised pipeline (Sec.~\ref{sec:circuit}) on bundled
  registers: product-state preparation
  $\lvert g\rangle\otimes\lvert\psi_0\rangle$, $\mathrm{QFT}^{\dagger}$
  on the $p$ register, Strang-split steps alternating the Bell wave
  circuit $e^{-i\Delta t H}$ with the diagonal damping phases
  $e^{+i\Delta t\,\eta\Sigma}$ (half steps at the ends), closing
  $\mathrm{QFT}$, and recovery by measuring the $p$ register,
  post-selecting the slice at $p^{*}$, and rescaling classically by
  $e^{p^{*}}$.}
  \label{fig:circuits}
\end{figure}
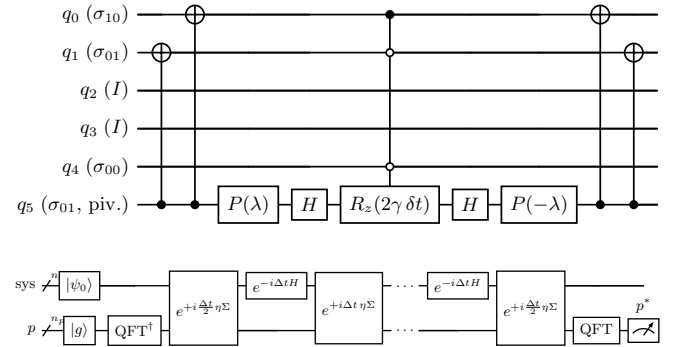

The circuit of Sato \emph{et al.}~\cite{sato2024hamiltonian}
exponentiates ladder-string terms (alphabet $\{I,\sigma_{01},
\sigma_{10}\}$) via a Bell-basis change followed by one multi-controlled
$R_z$; Hu, Jin, Liu, and Zhang~\cite{hu2024quantum} extended the
framework to non-unitary PDE dynamics via Schr\"odingerisation, and Jin,
Liu, and Yu~\cite{jin2024heat} built gate-level circuits for the heat
equation with physical boundaries.  Relative to these we add the
\emph{projector} extension: strings containing
$z=\lvert0\rangle\langle0\rvert$ and $o=\lvert1\rangle\langle1\rvert$
(block projectors of separated multi-field systems) enter as plain
$0$-/$1$-controls; the Bell rotation outside the projected subspace is
undone exactly, so the circuit implements
$e^{-i\,\delta t\,(cP+\mathrm{h.c.})}$ for projector-valued $P$.
Figure~\ref{fig:circuits}(a) shows the compiled circuit for a
representative projector-containing term of the separated 2D wave
Hamiltonian, and Fig.~\ref{fig:circuits}(b) the full Schr\"odingerised
pipeline it slots into.

Due diligence (full-text check of both papers): neither
Ref.~\cite{hu2024quantum} nor Ref.~\cite{jin2024heat} circuitises
projector-containing strings---both use only the
$\{I,\sigma_{01},\sigma_{10}\}$ alphabet.  We confirmed this at code
level in the published companion notebook of
Ref.~\cite{hu2024quantum}: its term circuit ($W_j$) implements the
contiguous ladder pattern with rotations controlled uniformly on all
lower qubits (an all-or-nothing $X$-conjugation flag serving the
periodic corner terms), with no per-qubit $0$-/$1$-control selection
(the nilpotent shift needs no
projector guards in scalar single-field problems, and
Ref.~\cite{jin2024heat} routes boundary data into source terms precisely
to avoid non-Toeplitz structure).  Ref.~\cite{sato2025nonconservative}
comes closest: its symbolic operator algebra over
$\{I,\sigma_{00},\sigma_{11},\sigma_{01},\sigma_{10}\}$ produces
projector factors from boundary-truncated shifts and piecewise-constant
coefficients, and poly-depth Trotterisation is asserted by citation to
Ref.~\cite{sato2024hamiltonian}; but no gate-level compilation rule for
the projector factors and no projector-term circuit is given.  A caveat
we state explicitly: the
abstract class $\lvert a\rangle\langle b\rvert + \lvert b\rangle\langle
a\rvert$ over basis states in Lemma~1 of
Refs.~\cite{sato2024hamiltonian,hu2024quantum} implicitly \emph{contains}
Hermitian-paired projector strings as an existence statement; our
contribution is the explicit alphabet extension and compilation rule
(projector factor $\to$ $0$-/$1$-control) and its application to
multi-field block-coupled systems, not a new diagonalisation principle.
The extension is backward compatible.  Unitary-level validation: the
error is pure per-step Trotter error $0.50\,\delta t^{2}$ with no phase
offset.  This unlocks the separated 2D wave Hamiltonian
(\texttt{zm}/\texttt{mz} block couplings) needed for per-direction
absorbing layers.

\subsection{The Schr\"odingerised circuit}
\label{sec:circuit}

The full pipeline is
\begin{equation}
\begin{split}
  \mathrm{init}(g\otimes\psi_0)
  \;\to\; \mathrm{QFT}^{\dagger}_p
  \;\to\; [\text{Trotter steps}] \\
  \;\to\; \mathrm{QFT}_p
  \;\to\; \text{slice at } p^{*}.
\end{split}
  \label{eq:pipeline}
\end{equation}
Each step (diagonal $H_1$) factorises into the wave circuit (the Bell
construction of Sec.~\ref{sec:projector} on the system register) and a
diagonal damping phase with entries $e^{+i\,\delta t\,\eta_k\sigma_j}$.
Both Trotter levels are symmetrised (Strang outer; palindromic term order
inner), with measured clean second order: errors
$4\times10^{-2}\to10^{-2}\to2.7\times10^{-3}$ at $50/100/200$ steps on 17
qubits.

\emph{Fitted orders} [1D benchmark, error versus the exact
Schr\"odingerised evolution to isolate the splitting;
Fig.~\ref{fig:convergence}(d)]: slope $-1.02$ (order~1), $-2.01$ (order~2
with both levels symmetrised); against the full matrix-exponential
reference the order-2 error saturates at the $p$-discretisation floor, as
expected.  \emph{Warping-profile order} [recovery error versus $n_p$,
fixed 1D PML case; Fig.~\ref{fig:convergence}(c)]: $e^{-|p|}$ slope
${\sim}{-1}$ in $\Delta p$ ($4.1\times10^{-2}\to6.3\times10^{-4}$ over
$n_p=6$--$11$); $C^{1}$ cubic ${\sim}{-2}$
($1.3\times10^{-2}\to2.4\times10^{-6}$), confirming Remark~4.9 of
Ref.~\cite{jin2024recovery} in this setting.

\subsection{End-to-end demonstrations}
\label{sec:demos}

\begin{figure*}
  \includegraphics[width=\textwidth]{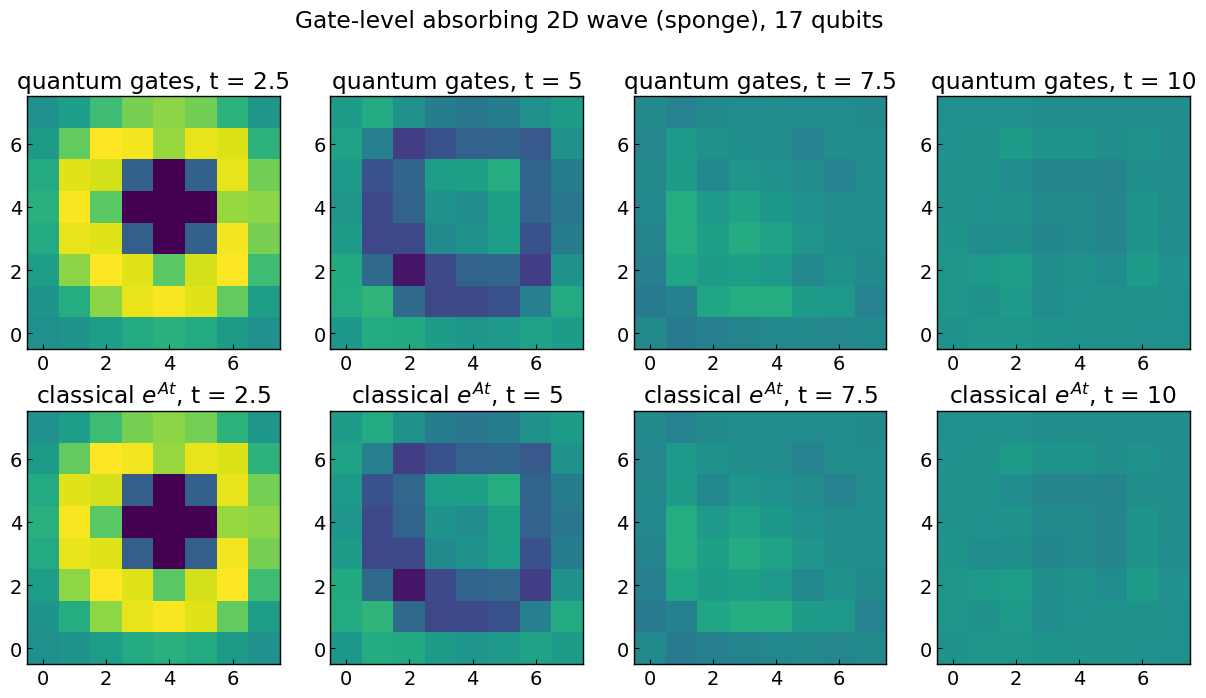}
  \caption{Gate-level absorbing 2D wave evolution (sponge layer,
  $8\times8$ grid, 17 qubits $=$ 6 grid $+$ 2 block $+$ 9 $p$-register).
  Top row: snapshots of the recovered field from the Trotterised quantum
  circuit at $t=2.5,5,7.5,10$.  Bottom row: the exact non-unitary
  reference $e^{At}\psi_0$.  The expanding pulse is absorbed in the
  boundary layer with no visible reflection; snapshot errors are at the
  $10^{-3}$ level (Sec.~\ref{sec:demos}).}
  \label{fig:gate17}
\end{figure*}

\begin{figure*}
  \includegraphics[width=\textwidth]{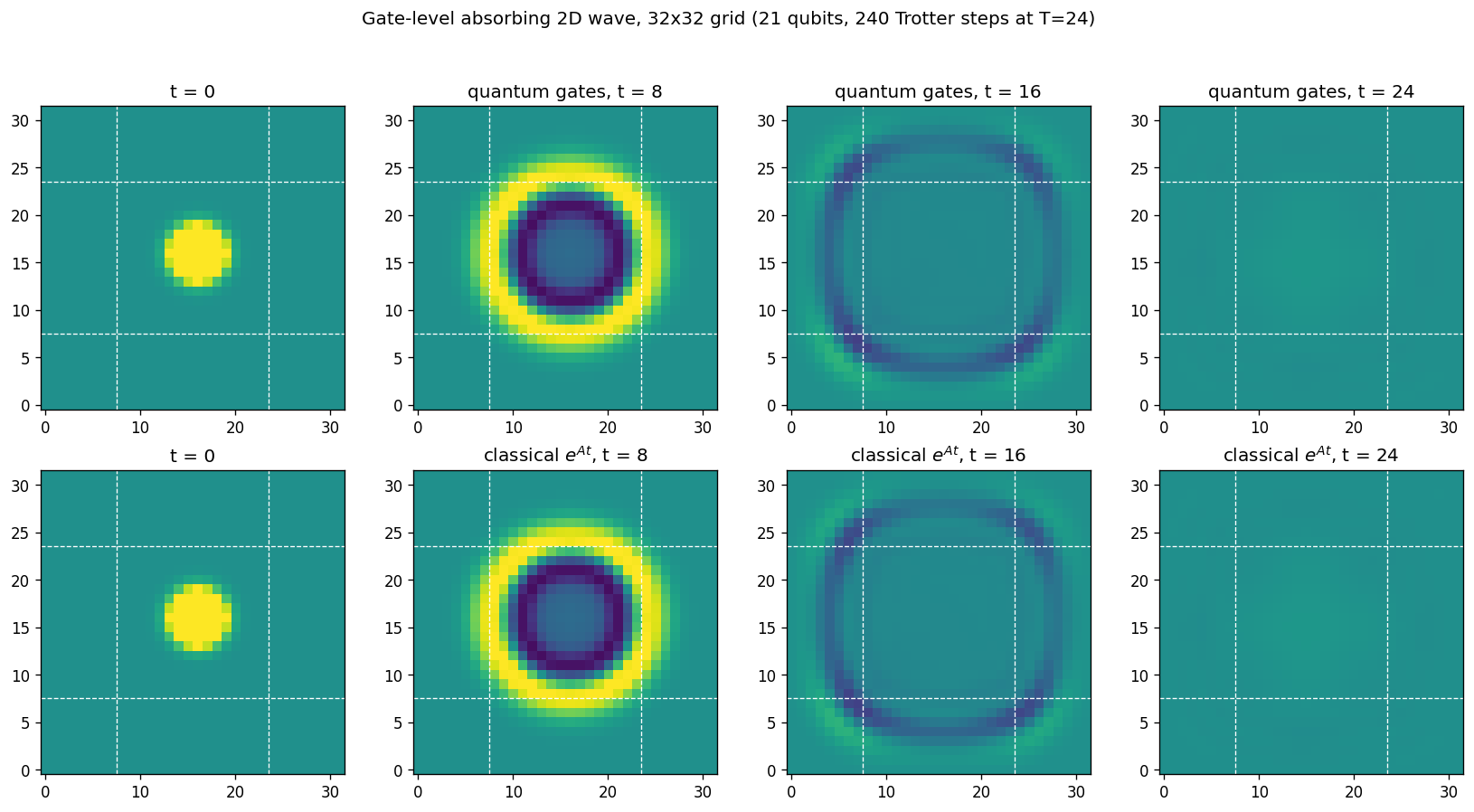}
  \caption{Gate-level absorbing 2D wave evolution at scale: $32\times32$
  grid, 21 qubits (12 system $+$ 9 $p$-register), up to 240 second-order
  Trotter steps (983 native operations per step).  Top row: recovered
  field from the quantum circuit at $t=0,8,16,24$ (dashed lines mark the
  absorbing-layer strips).  Bottom row: exact reference $e^{At}\psi_0$.
  The outgoing ring is absorbed across the layer; the total energy decays
  from $0.97$ to $1.25\times10^{-3}$, and the recovered field matches the
  reference to $7.7\times10^{-3}$--$1.7\times10^{-2}$.}
  \label{fig:gate21}
\end{figure*}

\begin{itemize}
\item 1D CPML, 14 qubits (6 system $+$ 8 $p$), 60 steps, $T=30$:
  recovered solution within $7\times10^{-4}$ of the exact non-unitary
  evolution.
\item 2D sponge, $8\times8$ grid, 17 qubits, $T=10$:
  $2.7\times10^{-3}$ at 200 steps (Fig.~\ref{fig:gate17}).
\item 2D sponge, $32\times32$ grid, 21 qubits (12 system $+$ 9 $p$), up
  to 240 steps (983 native operations per step): errors
  $7.7\times10^{-3}$ / $1.7\times10^{-2}$ / $1.4\times10^{-2}$ at
  $T=8/16/24$; energy decays from $0.97$ to $1.25\times10^{-3}$
  (absorption complete); nine minutes of statevector simulation
  (Fig.~\ref{fig:gate21}).
\end{itemize}
Qubit accounting: the absorbing overhead is $+1$ (sponge) or $+2$ (CPML)
block qubits plus $n_p\approx8$--$10$ for the $p$ register; $n_p$ depends
only logarithmically on $T$, the damping rate, and the target accuracy.

\section{Complexity}
\label{sec:complexity}

\begin{figure}
  \includegraphics[width=\columnwidth]{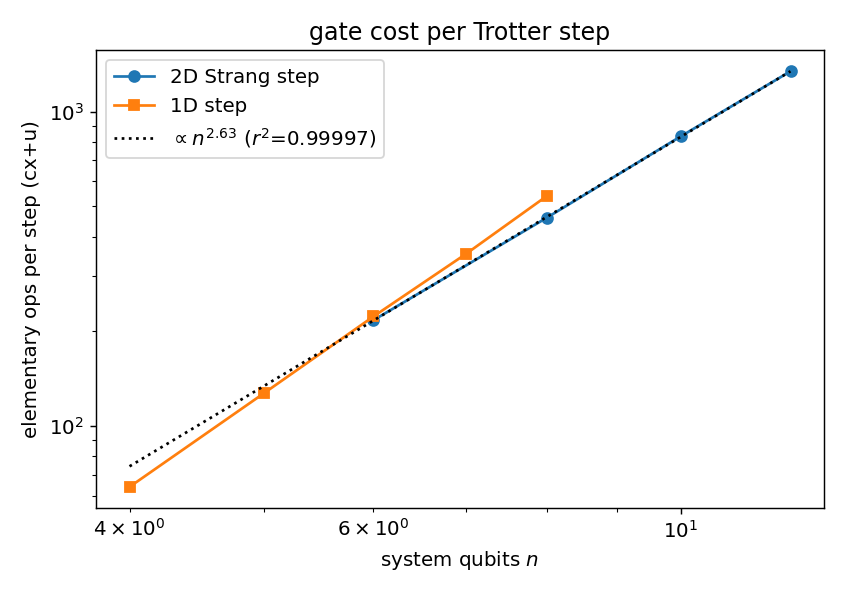}
  \caption{Gate cost per Trotter step after transpilation to the
  $\{\mathrm{CX},u\}$ basis, versus the number of system qubits $n$: 1D
  steps (squares) and 2D Strang steps (circles).  The dotted line is the
  power-law fit $\propto n^{2.63}$ ($r^{2}=0.99997$) to the 2D series;
  the term count itself is exactly linear in $n$ (Sec.~\ref{sec:complexity}).}
  \label{fig:gatecounts}
\end{figure}

Term counts inherit the QTT structure of the difference operators:
$2n_x+2$ terms in 1D and $4n_x+4$ in 2D ($n_x=n_y$), exactly linear in
the register size (fitted slopes $4.000/2.000$, intercepts $4/2$).
Transpiled to the $\{\mathrm{CX},u\}$ basis, one Strang wave step costs
$217/459/834/1343$ operations for $n_{\mathrm{sys}}=6/8/10/12$; a
power-law fit gives exponent $2.63$ with $r^{2}=0.99997$
(Fig.~\ref{fig:gatecounts})---convincingly polynomial, and in fact
asymptotically quadratic: the term count is linear in $n$ and the
ancilla-free V-chain multi-controlled-$R_z$ is linear in its control
count, so ops/step $=O(n^{2})$.  The constants are fixed by the
measurements.  The per-step cost in the
$\{\textsc{cx}, U\}$ basis is reproduced to $0.03\%$
($r^{2}=0.9999999$) by the quadratic
\begin{equation}
  G_{\mathrm{step}}(n) \;=\; 16.7\,n^{2} - 112.7\,n + 293,
  \label{eq:stepcost}
\end{equation}
whose structure is transparent: a Strang step exponentiates $2n$
Hermitian-paired terms ($4n_x+4$ at $n=2n_x+2$), each one Bell basis
change plus one multi-controlled $R_z$ whose ancilla-free V-chain cost
is \emph{exactly} $32k-31$ elementary operations for $k\ge8$ controls
(measured), linear in $k$; with the QTT ladder structure averaging
$\bar k\approx n/4$ controls per term, the leading coefficient is
$2n\cdot32\,\bar k/n^{2}\approx16.7$.  The total gate count is steps
$\times$ ops/step with steps $= T/\delta t$ and $\delta t$ set by the
second-order bound $\delta t = \sqrt{\varepsilon_{\mathrm{tr}}/(C_2
T)}$, where $C_2$ is the Strang error constant of the slope-$(-2)$ fits
of Fig.~\ref{fig:convergence}(d):
\begin{equation}
  G_{\mathrm{total}}
  \;=\; \sqrt{\frac{C_2\,T^{3}}{\varepsilon_{\mathrm{tr}}}}\;
        G_{\mathrm{step}}(n)
  \;=\; O\!\left(\frac{T^{3/2}}{\sqrt{\varepsilon_{\mathrm{tr}}}}\;
        n^{2}\right),
  \label{eq:totalcost}
\end{equation}
plus the damping diagonal per step and two $\mathrm{QFT}(n_p)$; the
power-law fit across the measured range reads $n^{2.63}$ because the
subleading terms of Eq.~\eqref{eq:stepcost} have not yet decayed at
$n\le12$.

\emph{Caveat (honest).}  The damping factor is a single native
\texttt{diagonal} instruction on the simulator, but a \emph{generic}
diagonal on $n$ qubits synthesises to $\Theta(2^{n})$ elementary gates:
we measure a CX count of exactly $2^{n}-2$ and totals within 15\% of the
$2^{n+1}-3$ synthesis bound.  For hardware the structure must be
exploited: the diagonal is $D_\eta\otimes\Sigma$ with $\eta$ linear in
the $p$ index and $\sigma$ a degree-$m$ polynomial profile on the layer
strips, so phase-polynomial constructions give $\mathrm{poly}(n)$
circuits, and the logic-minimisation synthesis of piecewise-constant
diagonal operators of Ref.~\cite{sato2025nonconservative}
(Espresso-compressed projector strings with matrix-product-state
coefficient oracles) provides an existing $\mathrm{poly}(n)$ toolbox
directly applicable to the strip-wise polynomial $\sigma$ profiles;
making this explicit is deferred to future work.

\section{Discussion and outlook}
\label{sec:discussion}

The $\lambda^{+}$ obstruction of Lemma~\ref{lem:threshold} is generic for
auxiliary-field absorbers---memory variables, auxiliary-differential-equation
formulations, split fields---because the injection/source asymmetry is
structural rather than an artefact of one discretisation.  The Lyapunov
symmetrizer is the corresponding general remedy, and its
quantum-implementable variants (block-encoding $S$, structured or banded
approximate symmetrizers) are open.  Natural next steps include a
hardware experiment in the style of Ref.~\cite{sato2024hamiltonian};
quantum-signal-processing or qubitisation in place of Trotter (and the
query-optimal Schr\"odingerisation formulation of
Ref.~\cite{jin2025queries}); CPML for Maxwell's equations
\cite{jin2023maxwell,ma2024maxwellcircuits}; and smoother warping
profiles (error-function or numerically optimal) beyond the $C^{1}$ cubic.

\medskip
\noindent\emph{Code and data availability.}  The implementation
(\texttt{lib/absorbing.py}, \texttt{lib/schrodingerisation.py}), all
experiment scripts and JSON result files
(\texttt{paper/experiments/}), and the executed demonstration notebook
are available in the accompanying repository at
\url{https://github.com/x-repos/qawave-cpml}.

\appendix

\section{Proof of Lemma~\ref{lem:threshold}}
\label{app:proof}

Fix a layer node $j$ with $\sigma_j>0$ and a stencil neighbour $i$ such
that the difference entry satisfies $|D_{ji}|=1/h$ (wave speed $c=1$;
the general case restores a factor $c$ in Eq.~\eqref{eq:lemmarate}).
Write $\beta=\sigma_j/\kappa_j+\alpha_j\ge0$ for the CFS decay rate,
$\gamma$ for the injection coefficient, and
$s=\sigma_j|D_{ji}|/(\kappa_j^{2}\gamma)$ for the memory-source
magnitude.  Let $G=\diag(g)>0$ be \emph{any} positive diagonal
similarity, and set
\begin{equation}
  u=\frac{g_{v_j}}{g_{\phi_j}}, \qquad
  t=\frac{g_{\phi_j}}{g_{w_i}}, \qquad
  \rho=u\,t=\frac{g_{v_j}}{g_{w_i}}.
  \label{eq:scalings}
\end{equation}
The transformed generator $\At = GAG^{-1}$ has unchanged diagonal, and
the three relevant coupling families scale as follows.
\begin{enumerate}
\item[(i)] \emph{Injection pair} $(v_j,\phi_j)$: entry $u\gamma$, reverse
  entry $0$, so $\Ht$ contains the principal $2\times2$ submatrix
  \begin{equation}
    \begin{bmatrix} 0 & u\gamma/2 \\ \bar u\bar\gamma/2 & -\beta
    \end{bmatrix}
    \;\Longrightarrow\;
    \lamax(\Ht) \ge f(|u|\gamma,\beta),
  \end{equation}
  with $f(x,\beta):=\bigl(\sqrt{\beta^{2}+x^{2}}-\beta\bigr)/2$.
\item[(ii)] \emph{Memory-source pair} $(w_i,\phi_j)$: entry $t\,s$
  analogously, giving $\lamax(\Ht)\ge f(|t|s,\beta)$.
\item[(iii)] \emph{Physical wave pair} $(v_j,w_i)$: elementwise,
  $\At[v,w]=-iD_{ji}\,\rho$ and $\At[w,v]=+iD_{ji}/\rho$ (using that $D$
  is real and $D^{\dagger}=-D_{\mathrm{b}}$), so
  \begin{equation}
  \begin{split}
    \Ht[v_j,w_i] &= \frac{iD_{ji}}{2}\bigl(\rho^{-1}-\rho\bigr) \\
    \Longrightarrow\quad
    \lamax(\Ht) &\ge \frac{|D_{ji}|}{2}\,\bigl|\rho^{-1}-\rho\bigr|.
  \end{split}
  \end{equation}
\end{enumerate}
All three are lower bounds on $\lamax(\Ht)$ by Cauchy interlacing (the
eigenvalues of any principal submatrix interlace those of the full
Hermitian matrix).

\emph{Case analysis.}  If $|\rho|\le1/2$ then
$|\rho^{-1}-\rho|\ge3/2$, so (iii) gives $\lamax(\Ht)\ge3/(4h)$.
Otherwise $|\rho|>1/2$ implies
\begin{equation}
  |u\gamma|\cdot|t\,s| \;=\; |\rho|\,\gamma s \;>\; \frac{\gamma s}{2}
  \;=\; \frac{\sigma_j|D_{ji}|}{2\kappa_j^{2}},
\end{equation}
hence
$\max(|u|\gamma,\,|t|s)\ge\sqrt{\sigma_j/(2h\kappa_j^{2})}$ and
(i)--(ii) give
$\lamax(\Ht)\ge f\bigl(\sqrt{\sigma_j/(2h\kappa_j^{2})},\,\beta\bigr)$.
Therefore, for every positive diagonal scaling,
\begin{equation}
  \lamax(\Ht) \;\ge\; \min\left\{\frac{3}{4h},\;
  \frac{\sqrt{\beta^{2}+\sigma_j/(2h\kappa_j^{2})}-\beta}{2}\right\},
  \label{eq:appbound}
\end{equation}
which is $\Theta\bigl(\sqrt{\sigma_j/h}\bigr)$ as $h\to0$ (the second
branch is the minimum once $\sigma_j/(2h)\gg\beta^{2}$, i.e.\ for any
fixed profile and small $h$).  The bound is independent of $\gamma$ and
of the scaling $G$.  The explicit balanced choice
$\gamma^{*}=\sqrt{2\sigma_{\max}/h}$ attains the same rate from above:
measured $\lambda^{+}=0.68$--$0.97$ across our configurations, versus the
lower bound $0.18$ at the $8\times8$, $\sigma_{\max}=0.5$ parameters---the
bound is a rate, not a sharp constant. \hfill$\blacksquare$

\medskip
\emph{Consistency check.}  The collapsed 1D form
(Sec.~\ref{sec:1d}) has no injection pair: the memory is eliminated and
the damping is diagonal, so the lemma does not apply---and indeed
$H_1=-\diag(\sigma)\preceq0$ there.

\input{sections/appB}
\input{sections/appC}
\input{sections/appD}
\input{sections/appE}

\bibliographystyle{apsrev4-2}
\bibliography{refs}

\end{document}

%% file: sections/cpml_vs_sponge.tex

\emph{CPML versus sponge at matched resources.}  A matched-resource
sweep quantifies this tradeoff: both absorbers on the same $32\times32$
grid with equal layer widths $n_{\mathrm{pml}}\in\{4,8,12\}$ and
default parameters, each measured against one shared $4\times$ enlarged
($128\times128$) hard-wall reference by the gold-standard protocol of
Sec.~\ref{sec:2d}, the reflection being the maximum interior error over
$T\in\{8,16,24,32,40\}$ (Table~\ref{cmp:tab:matched}).  At equal layer
budget the CPML reaches its discrete wall-return level of
$2.7\times10^{-4}$ by $n_{\mathrm{pml}}=8$ [consistent with the
$2.8\times10^{-4}$ gold-standard value of Sec.~\ref{sec:2d}], while the
sponge stays at the percent level, improving only from
$6.2\times10^{-2}$ to $3.6\times10^{-2}$ across the sweep---a quality
gap of a factor of $170$ at $n_{\mathrm{pml}}=8$ ($130$ at $12$).  The
quantum-side cost inverts the ranking.  The sponge has $\lambda^{+}=0$
exactly, so its recovery post-selects at the $T$-independent
$e^{2p^{*}}\approx1.2$ (default slice $p^{*}\approx0.11$); the CPML
pays $e^{2\lambda^{+}T}$, with $\lambda^{+}=0.93$--$1.54$ tracking the
$\Theta(\sqrt{\sigma_{\max}c/h})$ rate of Lemma~\ref{lem:threshold} to
within $5\%$---already $7\times10^{7}$ at $T=8$ and $1.6\times10^{39}$
at $T=40$ for $n_{\mathrm{pml}}=8$.  Neither absorber dominates at
matched resources: the sponge trades two orders of magnitude in
reflection for an essentially free recovery, and the CPML trades an
exponentially growing post-selection penalty for matched-layer
accuracy---precisely the penalty that the symmetrizer of
Sec.~\ref{sec:symmetrizer} replaces by a $T$-independent conditioning
factor.

\begin{table}[tb]
\caption{CPML versus sponge at matched resources: $32\times32$ grid,
equal layer width $n_{\mathrm{pml}}$, default parameters [$m=2$,
$R_0=10^{-3}$, Collino--Tsogka amplitudes
$\sigma_{\max}=2.59/1.30/0.86$ at $n_{\mathrm{pml}}=4/8/12$].
Reflection is the gold-standard interior error against a single shared
$4\times$ enlarged ($128\times128$) hard-wall reference, maximised over
$T\in\{8,16,24,32,40\}$ ($\lVert z_0\rVert=1$).
$\lambda^{+}=\lamax^{+}(H_1)$ sets the CPML post-selection penalty
$e^{2\lambda^{+}T}$ (quoted at $T=40$) and tracks
$\sqrt{\sigma_{\max}}$ to within $5\%$; the sponge has
$\lambda^{+}=0$, post-selecting at the $T$- and
$n_{\mathrm{pml}}$-independent $e^{2p^{*}}=1.2$ for the default
recovery slice $p^{*}=0.11$.}
\label{cmp:tab:matched}
\begin{ruledtabular}
\begin{tabular}{lllll}
 & \multicolumn{2}{c}{Reflection (max over $T$)} &
   \multicolumn{2}{c}{CPML recovery cost} \\
\cline{2-3}\cline{4-5}
$n_{\mathrm{pml}}$ & CPML & sponge & $\lambda^{+}(H_1)$ &
$e^{2\lambda^{+}T}$, $T{=}40$ \\
\colrule
4  & $4.2\times10^{-3}$ & $6.2\times10^{-2}$ & 1.54 & $2.2\times10^{53}$ \\
8  & $2.7\times10^{-4}$ & $4.7\times10^{-2}$ & 1.13 & $1.6\times10^{39}$ \\
12 & $2.8\times10^{-4}$ & $3.6\times10^{-2}$ & 0.93 & $1.5\times10^{32}$ \\
\end{tabular}
\end{ruledtabular}
\end{table}

%% file: sections/appB.tex
\section{The convolutional PML in the Hamiltonian-simulation variables}
\label{appB:cpml}

This appendix expands the one-line statements of
Secs.~\ref{sec:cpml}, \ref{sec:1d}, and \ref{sec:2d}: the causal kernel
and memory ODE from the CFS stretch
(Appendix~\ref{appB:kernel}), the one-dimensional memory generator and
the frequency-domain collapse (Appendix~\ref{appB:1d}), the proof that
the collapse is exact (Appendix~\ref{appB:collapse}), the
two-dimensional eight-block generator
(Appendix~\ref{appB:2d}), the staggered profile-sampling rule
(Appendix~\ref{appB:stagger}), and the profile formulas
(Appendix~\ref{appB:profiles}).  Throughout, $D$ is the
forward-difference matrix and $D^{\dagger}=-D_{\mathrm{b}}$ its negative
backward counterpart (Sec.~\ref{sec:wave}); all operators are exactly
those constructed in the accompanying code.

\subsection{From the CFS stretch to the memory ODE}
\label{appB:kernel}

We fix the Fourier convention
$\hat f(\omega)=\int dt\,e^{-i\omega t}f(t)$,
$f(t)=(2\pi)^{-1}\int d\omega\,e^{+i\omega t}\hat f(\omega)$, so that
$\partial_t\mapsto i\omega$ and the causal exponential has the transform
$\int_0^{\infty}dt\,e^{-\beta t}e^{-i\omega t}=(\beta+i\omega)^{-1}$ for
$\beta>0$.  A PML stretches the coordinate into the complex plane,
$\tilde x=\int_0^x s(x',\omega)\,dx'$, which in the frequency domain
replaces every spatial derivative by
$\partial_x\mapsto s(\omega)^{-1}\partial_x$ inside the layer.  For the
CFS factor of Eq.~\eqref{eq:cfs}, partial fractions in the variable
$i\omega$ give
\begin{equation}
  \frac{1}{s(\omega)}
  \;=\; \frac{\alpha+i\omega}{\kappa(\alpha+i\omega)+\sigma}
  \;=\; \frac{1}{\kappa} \;+\; \hat\zeta(\omega),
  \label{appB:eq:invs}
\end{equation}
with
\begin{equation}
  \hat\zeta(\omega) \;=\; -\frac{\sigma}{\kappa^{2}}\,
  \frac{1}{\beta+i\omega},
  \qquad
  \beta \;:=\; \frac{\sigma}{\kappa}+\alpha
  \label{appB:eq:zetahat}
\end{equation}
the CFS decay rate: indeed
$1/s-1/\kappa=(\kappa-s)/(\kappa s)$ with
$\kappa-s=-\sigma/(\alpha+i\omega)$ and
$\kappa s=\kappa^{2}(\beta+i\omega)/(\alpha+i\omega)$.  The single pole
at $i\omega=-\beta$ inverts, by the causal transform above, to the
exponential kernel of Eq.~\eqref{eq:kernel},
$\zeta(t)=-(\sigma/\kappa^{2})\,e^{-\beta t}\theta(t)$, and frequency
multiplication is time convolution:
$s^{-1}\partial_x f = \kappa^{-1}\partial_x f + \zeta * \partial_x f$.

The convolution becomes local in time once it is named.  Let
$\phi:=\zeta*\partial_x f$.  Multiplying Eq.~\eqref{appB:eq:invs} by
$(\beta+i\omega)$ gives the polynomial identity
$(\beta+i\omega)\hat\zeta=-\sigma/\kappa^{2}$, hence
$i\omega\,\hat\phi=-\beta\,\hat\phi
-(\sigma/\kappa^{2})\,\widehat{\partial_x f}$, which transforms back to
the memory ODE of Eq.~\eqref{eq:memoryode}.  Equivalently, in the time
domain,
$\partial_t\phi=\zeta(0^{+})\,\partial_x f+\zeta'*\partial_x f$ with
$\zeta(0^{+})=-\sigma/\kappa^{2}$ and $\zeta'=-\beta\zeta$.  One
first-order ODE per stretched derivative thus carries the convolution
exactly; this is the recursive-convolution CPML of Roden and
Gedney~\cite{roden2000cpml}.

\subsection{One dimension: memory generator and collapse}
\label{appB:1d}

In the variables of Sec.~\ref{sec:wave} the undamped 1D system is
$\dot v=-iDw$, $\dot w=-iD^{\dagger}v$ [Eq.~\eqref{eq:waveH}].  Each
equation contains exactly one spatial derivative, so the stretch
generates one memory field per physical field.  Sampling the profiles
per field (Appendix~\ref{appB:stagger}) defines the diagonal matrices
$\Sigma_v=\diag(\sigma_{v,j})$, $K_v=\diag(\kappa_{v,j})$,
$B_v=\diag(\sigma_{v,j}/\kappa_{v,j}+\alpha_{v,j})$, and the layer
projector $\Pi_v=\diag(\mathbf{1}[\sigma_{v,j}>0])$, and likewise for
$w$.  Applying the stretched-derivative rule to $-iDw$ and carrying the
convolution by the rescaled memory field
$\phi_v:=\gamma^{-1}\,\zeta_v*(-iDw)$---where $\gamma>0$ is a free
constant rescaling under which the physical trajectory is invariant,
set to the balanced value $\gamma=\sqrt{2\sigma_{\max}/h}$ of
Appendix~\ref{app:proof}---gives, by Eq.~\eqref{eq:memoryode},
\begin{equation}
  \dot v = -iK_v^{-1}Dw + \gamma\phi_v,
  \qquad
  \dot\phi_v = -B_v\phi_v + \frac{i}{\gamma}\Sigma_vK_v^{-2}Dw,
  \label{appB:eq:stretched}
\end{equation}
and the analogous pair for $(w,\phi_w)$ with $D\to D^{\dagger}$,
$v\leftrightarrow w$.  The source of $\phi_v$ carries the factor
$\Sigma_v$, so $\phi_v$ vanishes identically outside the layer strips
for memory-free data; masking the injection column by $\Pi_v$ therefore
changes no such trajectory while removing the spurious $O(\gamma)$
interior entries that an unmasked column would contribute to the
Hermitian part $H_1$.  On the four-block state
$[v;\,w;\,\phi_v;\,\phi_w]$ (two extra block qubits) the generator is
\begin{equation}
  A_{\mathrm{mem}} =
  \begin{bmatrix}
    0 & -iK_v^{-1}D & \gamma\Pi_v & 0\\[2pt]
    -iK_w^{-1}D^{\dagger} & 0 & 0 & \gamma\Pi_w\\[2pt]
    0 & \tfrac{i}{\gamma}\Sigma_vK_v^{-2}D & -B_v & 0\\[2pt]
    \tfrac{i}{\gamma}\Sigma_wK_w^{-2}D^{\dagger} & 0 & 0 & -B_w
  \end{bmatrix}.
  \label{appB:eq:gen1d}
\end{equation}

\emph{Collapse for $\kappa=1$, $\alpha=0$.}  In this case the stretch
degenerates: $s(\omega)=(i\omega+\sigma)/(i\omega)$, so the stretched
$v$ equation, $i\omega\hat v=s_v(\omega)^{-1}(-iD\hat w)$ per Fourier
mode, multiplies out to
\begin{equation}
  (i\omega+\sigma_v)\,\hat v \;=\; -iD\hat w,
  \label{appB:eq:freqcollapse}
\end{equation}
because the symbol $s(\omega)\,i\omega=i\omega+\sigma$ is a
\emph{first-degree polynomial} in $i\omega$.  Transforming back,
$\dot v=-iDw-\sigma_v v$, and likewise
$\dot w=-iD^{\dagger}v-\sigma_w w$: the convolution collapses to the
local damping of Eq.~\eqref{eq:1dcollapse},
$A=-iH-\diag(\sigma_v,\sigma_w)$, with no memory fields at all.

\subsection{Proof of the exact collapse}
\label{appB:collapse}

The frequency-domain argument above is formal (it commutes the stretch
with the time-domain initial-value problem); the following statement
makes the equivalence exact at the semi-discrete level.

\begin{lemma}[exact collapse]
\label{appB:lem:collapse}
Let $\kappa\equiv1$ and $\alpha\equiv0$ in the generator of
Eq.~\eqref{appB:eq:gen1d}, so that $B_v=\Sigma_v$ and
$B_w=\Sigma_w$, and define the defects
\begin{equation}
  e_v := \phi_v + \gamma^{-1}\Sigma_v v,
  \quad
  e_w := \phi_w + \gamma^{-1}\Sigma_w w.
  \label{appB:eq:defect}
\end{equation}
Then $\dot e_v=\dot e_w=0$ along every solution of
$\dot\Psi=A_{\mathrm{mem}}\Psi$.  Consequently a memory-form trajectory
projects onto a trajectory of the collapsed generator
Eq.~\eqref{eq:1dcollapse} if and only if $e_v(0)=e_w(0)=0$; for the
standard embedding $\phi_v(0)=\phi_w(0)=0$ this is the condition
$\Sigma_v v(0)=0$ and $\Sigma_w w(0)=0$, i.e.\ initial data supported
outside the layer.
\end{lemma}

\begin{lemproof}
Using both rows of Eq.~\eqref{appB:eq:stretched} with $B_v=\Sigma_v$,
\begin{equation*}
\begin{split}
  \dot e_v
  &= \Bigl[-\Sigma_v\phi_v+\tfrac{i}{\gamma}\Sigma_vDw\Bigr]
  + \gamma^{-1}\Sigma_v\bigl[-iDw+\gamma\Pi_v\phi_v\bigr]\\
  &= (\Sigma_v\Pi_v-\Sigma_v)\,\phi_v \;=\; 0,
\end{split}
\end{equation*}
since $\Sigma_v\Pi_v=\Sigma_v$ (the projector is the support indicator
of $\Sigma_v$); identically for $e_w$.  Equivalently, the block-row
operator
$E=\bigl[\begin{smallmatrix}
\gamma^{-1}\Sigma_v & 0 & I & 0\\
0 & \gamma^{-1}\Sigma_w & 0 & I
\end{smallmatrix}\bigr]$
annihilates the generator from the left, $EA_{\mathrm{mem}}=0$
(verified to round-off in our implementation).  On the invariant slice
$e\equiv0$ the memory ansatz $\phi_v=-\gamma^{-1}\Sigma_v v$ of
Sec.~\ref{sec:1d} holds at all times, the injection term becomes
$\gamma\Pi_v\phi_v=-\Sigma_v v$, and $(v,w)$ obeys
Eq.~\eqref{eq:1dcollapse}; conversely, every collapsed trajectory lifts
to the memory form through the ansatz.
\end{lemproof}

When $e(0)\neq0$, the conserved defect acts on the physical equations
as the static in-layer forcing $\gamma\Pi_ve_v(0)$,
$\gamma\Pi_we_w(0)$, so the discrepancy between the two forms scales
with the layer content $\lVert\Sigma_vv(0)\rVert$,
$\lVert\Sigma_ww(0)\rVert$ of the initial data---and vanishes exactly
for data supported outside the layer.  Numerically, propagating the
same interior pulse with Eq.~\eqref{appB:eq:gen1d} and with
Eq.~\eqref{eq:1dcollapse} agrees to $6\times10^{-13}$
(Sec.~\ref{sec:1d}).

\emph{Remark (the roles of $\alpha$ and $\kappa$).}  The cancellation
uses only $\alpha\equiv0$: for a graded $\kappa$ the modified defect
$e_v=\phi_v+\gamma^{-1}\Sigma_vK_v^{-1}v$ is conserved by the same
computation, and the convolution collapses to the local form
$\dot v=-iK_v^{-1}Dw-K_v^{-1}\Sigma_vv$ (both statements again verified
to round-off).  For $\alpha>0$, instead, $\dot e_v=-\diag(\alpha_v)\phi_v
\neq0$---equivalently, $s(\omega)\,i\omega$ is no longer polynomial in
$i\omega$---and the memory fields are irreducible: the four-block form
is then the genuine content of the CFS layer.

\subsection{Two dimensions: the eight-block generator}
\label{appB:2d}

In 2D the damping acts per direction: the $x$ stretch replaces every
$\partial_x$ and the $y$ stretch every $\partial_y$.  On the
separated-block state of Sec.~\ref{sec:wave} the undamped equations are
$\dot v=-iD_xw_x-iD_yw_y$, $\dot w_x=-iD_x^{\dagger}v$,
$\dot w_y=-iD_y^{\dagger}v$, with $D_x=D\otimes I_{N_y}$ and
$D_y=I_{N_x}\otimes D$ on the flattened $N_xN_y$ grid register ($x$
register more significant).  The $v$ equation contains one derivative
of each direction and acquires two memory fields $\phi_{vx}$,
$\phi_{vy}$; each $w$ equation contains a single derivative and
acquires one, $\phi_x$ for $w_x$ and $\phi_y$ for $w_y$.  The four
memory fields plus the three physical fields are padded by one zero
block to the eight blocks of Eq.~\eqref{eq:blocks} (three block
qubits).

Because $\sigma_x$ depends on $x$ only, its diagonal samples factorise
on the grid register:
$\Sigma_v^{x}=S_x\otimes I_{N_y}$ and
$\Sigma_w^{x}=S_x^{\mathrm{st}}\otimes I_{N_y}$, where
$S_x=\diag\,\sigma(x_j)$ holds the nodal samples and
$S_x^{\mathrm{st}}=\diag\,\sigma(x_{j-1/2})$ the half-cell samples
(Appendix~\ref{appB:stagger}); symmetrically
$\Sigma_v^{y}=I_{N_x}\otimes S_y$ and
$\Sigma_w^{y}=I_{N_x}\otimes S_y^{\mathrm{st}}$.  With $\kappa\equiv1$
(our 2D configurations), the decay rates are
$B_v^{x}=\Sigma_v^{x}+\diag\alpha_x(x_j)$ etc., and the masks
$\Pi_v^{x}=\diag(\mathbf{1}[\Sigma_v^{x}>0])$, \dots\ are the
layer-strip projectors.  Repeating the derivation of
Appendix~\ref{appB:1d} once per (field, stretched direction) pair gives
the generator, in the block ordering of Eq.~\eqref{eq:blocks},
\begin{widetext}
\begin{equation}
  A \;=\;
  \begin{bmatrix}
    0 & -iD_x & -iD_y & \gamma\Pi_v^{x} & \gamma\Pi_v^{y} & 0 & 0 & 0\\[3pt]
    -iD_x^{\dagger} & 0 & 0 & 0 & 0 & \gamma\Pi_w^{x} & 0 & 0\\[3pt]
    -iD_y^{\dagger} & 0 & 0 & 0 & 0 & 0 & \gamma\Pi_w^{y} & 0\\[3pt]
    0 & \frac{i}{\gamma}\Sigma_v^{x}D_x & 0 & -B_v^{x} & 0 & 0 & 0 & 0\\[3pt]
    0 & 0 & \frac{i}{\gamma}\Sigma_v^{y}D_y & 0 & -B_v^{y} & 0 & 0 & 0\\[3pt]
    \frac{i}{\gamma}\Sigma_w^{x}D_x^{\dagger} & 0 & 0 & 0 & 0 & -B_w^{x} & 0 & 0\\[3pt]
    \frac{i}{\gamma}\Sigma_w^{y}D_y^{\dagger} & 0 & 0 & 0 & 0 & 0 & -B_w^{y} & 0\\[3pt]
    0 & 0 & 0 & 0 & 0 & 0 & 0 & 0
  \end{bmatrix}
  \quad\text{on}\quad
  \bigl[\,v;\; w_x;\; w_y;\; \phi_{vx};\; \phi_{vy};\;
        \phi_{x};\; \phi_{y};\; 0\,\bigr],
  \label{appB:eq:gen2d}
\end{equation}
\end{widetext}
which is, entry for entry, the matrix assembled by the accompanying
implementation.

\emph{Strip masking.}  The source of each memory field carries its own
profile factor ($\Sigma_v^{x}$ for $\phi_{vx}$, \dots), so the field
remains identically zero outside its layer strip for memory-free
initial data; the corresponding injection column is masked by the strip
projector, exactly as in 1D, so that $H_1$ carries no spurious
interior coupling.

\emph{Corners.}  In a corner cell both $\sigma_x>0$ and $\sigma_y>0$
hold simultaneously: both directions' memory fields are active there,
the same generator entries of Eq.~\eqref{appB:eq:gen2d} apply
unchanged, and no corner-specific terms arise.  This is the structural
advantage of the unsplit convolutional PML over split-field
formulations~\cite{roden2000cpml}.  The measured stability,
$\max\operatorname{Re}\operatorname{eig}(A)=0$ to round-off, is
reported in Sec.~\ref{sec:2d}.

\subsection{The staggered sampling rule}
\label{appB:stagger}

For the forward-difference convention, $(Dw)_j=(w_{j+1}-w_j)/h$.  The
component $v_j$ is updated through $(Dw)_j$ and represents the field at
the integer node $x_j=jh$; the component $w_j$ is updated through
$(D^{\dagger}v)_j\propto v_j-v_{j-1}$, a difference centered at the
half-cell point $x_{j-1/2}$, so the $w$ field lives on the staggered
grid (the backward convention shifts it to $x_{j+1/2}$).  The sampling
rule follows: \emph{every profile factor in an evolution equation is
sampled on the grid of the field that the equation updates},
\begin{equation}
  \sigma_{v,j}=\sigma(x_j),
  \qquad
  \sigma_{w,j}=\sigma(x_{j\mp1/2}),
  \label{appB:eq:staggerrule}
\end{equation}
with the upper (lower) sign for the forward (backward) convention,
and per direction in 2D (Table~\ref{appB:tab:stagger}): only the $x$
profile of the $w_x$ family and the $y$ profile of the $w_y$ family are
staggered; all other dependencies are nodal.

\begin{table}
\caption{Staggered sampling of the damping profiles in 2D
(forward-difference convention; the backward convention flips the
half-cell shifts).  Each profile is sampled on the grid of the field
its equation updates.}
\label{appB:tab:stagger}
\begin{ruledtabular}
\begin{tabular}{lll}
Field & Lives at & Profile samples \\
\colrule
$v$   & $(x_j,\,y_k)$         & $\sigma(x_j)$,\ $\sigma(y_k)$ \\
$w_x$ & $(x_{j-1/2},\,y_k)$   & $\sigma(x_{j-1/2})$ \\
$w_y$ & $(x_j,\,y_{k-1/2})$   & $\sigma(y_{k-1/2})$ \\
\end{tabular}
\end{ruledtabular}
\end{table}

\emph{Ablation.}  Sampling all profiles at the integer nodes
instead---the only change being the half-cell shift of the $w$
profiles---degrades the measured gold-standard reflection of
Sec.~\ref{sec:2d} by factors of 120--235.  At the reflection floors of
Fig.~\ref{fig:convergence}(a,b) the half-cell stagger is load-bearing,
not a refinement.

\subsection{Profile formulas}
\label{appB:profiles}

Within a layer of $n_{\mathrm{pml}}$ points (width
$L=n_{\mathrm{pml}}h$) the damping is polynomially graded in the depth
$d$ measured from the interface,
\begin{equation}
  \sigma(d) \;=\; \sigma_{\max}\,(d/L)^{m},
  \qquad 0\le d\le L,
  \label{appB:eq:sigma}
\end{equation}
and zero outside; the interface node itself is undamped,
$\sigma(0)=0$, the outermost node attains $\sigma_{\max}$, and
staggered samples beyond the wall are clipped to $d=L$.  The amplitude
follows from the normal-incidence round trip: a wave crossing the
layer, reflecting off the terminating hard wall, and crossing back
accumulates the attenuation
\begin{equation}
  R \;=\; \exp\Bigl[-\frac{2}{c}\int_0^{L}\sigma(x)\,dx\Bigr]
  \;=\; \exp\Bigl[-\frac{2\,\sigma_{\max}L}{(m+1)\,c}\Bigr],
  \label{appB:eq:roundtrip}
\end{equation}
so prescribing a design reflection $R_0$ gives the Collino--Tsogka
amplitude~\cite{collino2001pml}
\begin{equation}
  \sigma_{\max} \;=\; -\frac{(m+1)\,c\,\ln R_0}{2L},
  \label{appB:eq:sigmamax}
\end{equation}
used with the standard recipe $m=2$,
$R_0=10^{-3}$~\cite{komatitsch2007unsplit}.  The nominal $R_0$ is
realised only for layers of ${\gtrsim}8$ points; for thinner layers the
discrete transition error dominates (a four-point layer reflects at the
$0.5$--$2\%$ level for grid-scale wavelengths), consistent with the
saturation of Fig.~\ref{fig:convergence}(a).

When the CFS parameters are used they are graded with the same
geometry,
\begin{equation}
  \kappa(d) = 1 + (\kappa_{\max}-1)\,(d/L)^{m},
  \quad
  \alpha(d) = \alpha_{\max}\,(1-d/L),
  \label{appB:eq:grading}
\end{equation}
$\kappa$ rising from $1$ at the interface to $\kappa_{\max}$ at the
wall with the same polynomial as $\sigma$, and $\alpha$ tapering
linearly~\cite{komatitsch2007unsplit} from $\alpha_{\max}$ just inside
the interface to zero at the wall (and vanishing identically outside
the layer); in the implementation
$\alpha(d)=\alpha_{\max}[1-(\sigma(d)/\sigma_{\max})^{1/m}]$, which is
identical.  A nonzero $\alpha$ improves grazing-incidence and
evanescent absorption but unmatches the layer at low frequencies (the
stretch is matched only for $\omega\gg\alpha$), so the fixed-horizon
absorption benchmarks of Sec.~\ref{sec:2d} use $\alpha_{\max}=0$.
Finally, the memory rescaling defaults to
$\gamma=\sqrt{2\sigma_{\max}/h}$, the choice that balances the
injection and memory-source contributions to $\lamax(H_1)$
(Appendix~\ref{app:proof}).

%% file: sections/appC.tex
\section{Schr\"odingerisation: discrete formulas and recovery}
\label{appC:schro}

This appendix records the complete discrete recipe behind the summary of
Sec.~\ref{sec:schro}, with the exact sign, grid, and ordering conventions
of our implementation; these are the conventions validated end to end by
the demonstrations of Sec.~\ref{sec:demos}.

\subsection{Hermitian split and warped transform}
\label{appC:transform}

The semi-discrete absorbing dynamics $\dot z = Az$ is split as
\begin{equation}
  A = H_1 + iH_2, \qquad
  H_1 = \frac{A + A^{\dagger}}{2}, \qquad
  H_2 = \frac{A - A^{\dagger}}{2i},
  \label{appC:split}
\end{equation}
with the dissipative part $H_1$ and the Hamiltonian part $H_2$ both
Hermitian.  The warped phase-space transform of
Ref.~\cite{jin2023schrodingerisation} appends one auxiliary variable $p$
and sets $w(t,p) = e^{-p}z(t)$ for $p\ge0$ (extended to $p<0$ by the
profile $g$ of Sec.~\ref{appC:initial}).  The sign structure follows in
one line: along the exact solution, $\partial_t w = e^{-p}Az = Aw$ and
$\partial_p w = -e^{-p}z = -w$, so $H_1 w = -H_1\,\partial_p w$ and
\begin{equation}
  \partial_t w \;=\; -H_1\,\partial_p w \;+\; iH_2\,w,
  \label{appC:warped}
\end{equation}
a transport equation in $p$: the non-unitary part of $A$ has become
advection of the auxiliary variable.

\subsection{Discrete setup and per-mode evolution}
\label{appC:discrete}

The $p$ register holds $N_p = 2^{n_p}$ grid points on the periodic domain
$[-p_{\max},\,p_{\max})$,
\begin{equation}
  p_j = -p_{\max} + j\,\Delta p, \qquad
  \Delta p = \frac{2p_{\max}}{N_p},
  \label{appC:grid}
\end{equation}
$j = 0,\dots,N_p-1$, with dual frequencies $\eta_k = \pi k/p_{\max}$ for
$k = -N_p/2,\dots,N_p/2-1$ in FFT ordering [i.e.\
$\eta = 2\pi\times\texttt{fftfreq}(N_p,\Delta p)$].  We fix the Fourier
\emph{synthesis} convention $w(\cdot,p) = \sum_k \hat w_k\,e^{+i\eta_k p}$
(NumPy: $w = \mathrm{ifft}(\hat w)$ along the $p$ axis,
$\hat w = \mathrm{fft}(w)$).  Substituting
$\partial_p \to i\eta_k$ in Eq.~\eqref{appC:warped} decouples the modes
into the unitary evolutions of Eq.~\eqref{eq:permode},
\begin{equation}
\begin{split}
  \frac{d\hat w_k}{dt} &= -i\,(\eta_k H_1 - H_2)\,\hat w_k \\
  \Longrightarrow\quad
  \hat w_k(T) &= e^{-iT(\eta_k H_1 - H_2)}\,\hat w_k(0).
\end{split}
  \label{appC:permode}
\end{equation}
Stacking the modes, the whole evolution is one Hamiltonian simulation,
\begin{equation}
  H_{\mathrm{tot}} \;=\; D_\eta \otimes H_1 \;-\; I_p \otimes H_2,
  \qquad D_\eta = \diag(\eta_k),
  \label{appC:htot}
\end{equation}
acting on the tensor order ($p$ register, system register) with the $p$
register most significant: on the circuit, qubits $0,\dots,n_s-1$ carry
the system register (little-endian) and qubits $n_s,\dots,n_s+n_p-1$ the
$p$ register, so the flattened statevector index is $j\,2^{n_s} + s$.
The opening $\mathrm{QFT}^{\dagger}$ on the $p$ register maps the grid
amplitudes to $\hat w/\sqrt{N_p}$ (the unitarily normalised FFT),
$e^{-iTH_{\mathrm{tot}}}$ acts in the $\eta$ basis, and the closing
$\mathrm{QFT}$ returns to the $p$ grid [Fig.~\ref{fig:circuits}(b)].

\subsection{Initial data and the $C^{1}$ profile}
\label{appC:initial}

The initial condition $w(0,p_j) = g(p_j)\,z(0)$ is a product state across
the two registers,
\begin{equation}
  \lvert W(0)\rangle
  = \frac{\lvert g\rangle \otimes \lvert z_0\rangle}{N_0},
  \qquad
  N_0 = \lVert g\rVert_2\,\lVert z_0\rVert_2,
  \label{appC:product}
\end{equation}
with the normalisation $N_0$ recorded classically and undone at recovery.
The plain profile $g(p) = e^{-\lvert p\rvert}$ is kinked at $p=0$ and
limits the $p$ discretisation to $O(\Delta p)$; the $C^{1}$ cubic of
Ref.~\cite[Remark~4.9]{jin2024recovery} replaces it on $(-1,0)$,
\begin{equation}
  g(p) =
  \begin{cases}
    e^{-p}, & p \ge 0,\\[3pt]
    \Bigl(\dfrac{3}{e}-3\Bigr)p^{3}
      + \Bigl(\dfrac{4}{e}-5\Bigr)p^{2} - p + 1, & -1 < p < 0,\\[7pt]
    e^{p}, & p \le -1,
  \end{cases}
  \label{appC:profile}
\end{equation}
restoring $O(\Delta p^{2})$.  The junctions are $C^{1}$ by construction:
at $p=0^{-}$ the cubic gives $g=1$, $g'=-1$, matching $e^{-p}$; at
$p=-1^{+}$ it gives $g = g' = e^{-1}$, matching $e^{p}$.  Because
$g(p) = e^{-p}$ \emph{exactly} for $p\ge0$, the recovery formula below is
unchanged by the smoothing.  Measured on the 1D CPML benchmark of
Fig.~\ref{fig:convergence}(c) ($2^{5}$ spatial points,
$n_{\mathrm{pml}}=8$, $\sigma_{\max}=1$, $T=30$, $p_{\max}=18$): the
recovery error at the default slice falls from $4.1\times10^{-2}$ to
$6.3\times10^{-4}$ over $n_p = 6$--$11$ for the kinked profile and from
$1.3\times10^{-2}$ to $2.4\times10^{-6}$ for the cubic, with fitted
slopes of $\log_2(\mathrm{error})$ versus $n_p$ of $-1.2$ and $-2.6$
($-1.6$ and $-3.4$ on the pre-floor points $n_p\le9$)---at or above the
nominal first- and second-order rates, confirming the one-order upgrade
of Ref.~\cite{jin2024recovery} in this setting.

\subsection{Recovery: threshold, wraparound, plateau}
\label{appC:recovery}

The solution is recovered from a single grid slice,
\begin{equation}
  z(T) \;=\; e^{p^{*}}\,w(T,p^{*}),
  \qquad
  p^{*} \;\ge\; \lamax^{+}(H_1)\,T,
  \label{appC:recover}
\end{equation}
where $\lamax^{+} = \max\{0,\lamax(H_1)\}$ and the validity threshold on
$p^{*}$ is Theorem~3.1 of Ref.~\cite{jin2024recovery}.  When
$H_1\preceq0$ (the collapsed 1D CPML and the sponge) any $p^{*}>0$ is
admissible, and the implementation defaults to the slice three grid
points above $p=0$, i.e.\ $p^{*} = 3\Delta p$.  For indefinite $H_1$ (the
2D CPML memory form of Lemma~\ref{lem:threshold}) the slice must be
placed explicitly at $p^{*}\ge\lambda^{+}T$; below the threshold the
recovered state is $O(1)$ wrong (Fig.~\ref{fig:threshold}).

Two further discrete conditions guard the slice.  \emph{Wraparound}: in
Eq.~\eqref{appC:warped} an $H_1$-eigencomponent with eigenvalue $\lambda$
is advected in $p$ with velocity $\lambda$, so the most strongly damped
components travel a distance $\lambda^{-}T$ \emph{leftwards}, with
$\lambda^{-} = \lvert\lambda_{\min}(H_1)\rvert$ ($=\sigma_{\max}$ for the
diagonal absorbers).  The discrete $p$ domain is periodic, so
$p_{\max}$ must be wide enough that this leftward-transported mass does
not wrap through $p=-p_{\max}$ back into the recovery region.
\emph{Plateau}: since $e^{p}w(T,p) = z(T)$ for every admissible $p$, the
function $p\mapsto e^{p}\lVert w(T,p)\rVert$ must be flat across the
recovery window.  The implementation tabulates it over the first twelve
grid points right of $p=0$ and reports the relative peak-to-peak spread;
a flat plateau certifies that $p_{\max}$, $N_p$, and the profile are
adequate (our regression tests require a spread below $5\times10^{-2}$),
while a tilted or stepped plateau flags wraparound or an under-resolved
profile.

\subsection{Measurement interpretation}
\label{appC:measure}

The pipeline is unitary, so the final statevector is exactly
$W(T)/N_0$, holding $w_s(T,p_j)/N_0$ at index $j\,2^{n_s}+s$.  Measuring
the $p$ register returns outcome $p_j$ with probability
$\lVert w(T,p_j)\rVert^{2}/N_0^{2}$ and collapses the system register
onto $w(T,p_j)/\lVert w(T,p_j)\rVert = z(T)/\lVert z(T)\rVert$: every
slice in the admissible window yields the \emph{same} normalised state,
so post-selection may accept the whole plateau window rather than a
single outcome, summing the corresponding probabilities.  The success
probability of the slice at $p^{*}$ is, up to $p$-discretisation error,
\begin{equation}
  \Pr(p^{*})
  \;=\; \frac{e^{-2p^{*}}\,\lVert z(T)\rVert^{2}}{N_0^{2}}
  \;\propto\; e^{-2p^{*}},
  \label{appC:psuccess}
\end{equation}
which at the threshold slice $p^{*} = \lambda^{+}T$ is the
$e^{-2\lambda^{+}T}$ post-selection penalty of Sec.~\ref{sec:lemma}---the
cost that the Lyapunov symmetrizer of Sec.~\ref{sec:symmetrizer} replaces
by the $T$-independent factor $\kappa_2^{2}(S)$.  On the statevector
simulator used for the demonstrations of Sec.~\ref{sec:demos}, the slice
is read off rather than sampled: the final statevector is reshaped to an
$N_p\times2^{n_s}$ array, row $j$ at $p^{*}$ is extracted, and the result
is rescaled by $e^{p_j}N_0$.  This is deterministic and exact, so the
reported errors measure the Trotter and discretisation error only; on
hardware the same slice would additionally cost the factor
$e^{-2p^{*}}$ of Eq.~\eqref{appC:psuccess} in repetitions.

%% file: sections/appD.tex

\section{Validation protocol and convergence data}
\label{appD:validation}

This appendix records the classical validation protocol behind every
reflection number quoted in Sec.~\ref{sec:2d} and the raw convergence
data behind the fitted orders of Sec.~\ref{sec:circuit}
[Figs.~\ref{fig:convergence}(a--d)].

\subsection{Gold-standard reflection protocol}
\label{appD:protocol}

Measuring the residual reflection of an absorbing layer against the
absorbing dynamics itself is circular; the gold standard is the
open-domain solution.  The truncated domain has $N=128$ points (seven
grid qubits) with the CPML occupying $n_{\mathrm{pml}}$ points at each
end.  The initial data are a Gaussian velocity bump centred at $j=64$
with width $3$ cells and $w(0)=0$, normalised to
$\lVert z_0\rVert_2=1$.  The reference embeds the \emph{same} truncated
data centred in a $4\times$ hard-wall domain ($N_{\mathrm{ref}}=512$,
embedding offset $\Delta=(N_{\mathrm{ref}}-N)/2=192$) and evolves it
with the closed-domain unitary $e^{-iH_{\mathrm{ref}}T}$.  Both
evolutions are computed by dense matrix exponentials, so the comparison
contains no time-integration error.  The reported error is the
pointwise $\ell_2$ difference on the interior window,
\begin{equation}
  E(T)^{2} \;=\; \sum_{f\in\{v,\,w\}}\;
  \sum_{j=n_{\mathrm{pml}}}^{N-n_{\mathrm{pml}}-1}
  \bigl|\, f_j(T) - f^{\mathrm{ref}}_{j+\Delta}(T) \,\bigr|^{2}.
  \label{appD:eq:gold}
\end{equation}
Two checks make the protocol sound.  (i)~\emph{Window alignment}: at
$T=0$ the windowed difference in Eq.~\eqref{appD:eq:gold} evaluates to
exactly zero in floating point, pinning the embedding offset and the
index conventions.  (ii)~\emph{Causal validity}: with $c=1$, $h=1$ the
pulse meets the layer interface at depth $j=n_{\mathrm{pml}}$ and its
reflection is back at the centre by
$T\approx2(64-n_{\mathrm{pml}})\le120$ for $n_{\mathrm{pml}}\ge4$,
whereas the far walls of the reference domain are first felt inside the
comparison window only at $T\approx380$.  The plateau times
$T\in\{100,120,140\}$ therefore capture the complete round-trip
reflection of the truncated domain while the reference is still
uncontaminated.  The reported reflection is the maximum of $E(T)$ over
the three plateau times, and the relative plateau spread
[$(\max-\min)/\mathrm{mean}$] is the flatness diagnostic.  A reduced
instance of the same protocol ($N=64$ versus $256$, $T=20$ and $60$,
asserted $<10^{-3}$) is pinned in the unit-test suite
(Sec.~\ref{appD:tests}).

\subsection{Reflection versus layer width and design reflection}
\label{appD:reflection}

\begin{table}
\caption{Gold-standard true reflection (1D CPML, $N=128$ versus the
$N=512$ hard-wall reference, $m=2$): plateau error
$\max_{T\in\{100,120,140\}}E(T)$ and relative plateau spread.  Upper
block: layer-width sweep at $R_0=10^{-3}$; lower block: design-reflection
sweep at $n_{\mathrm{pml}}=12$.}
\label{appD:tab:reflection}
\begin{ruledtabular}
\begin{tabular}{llll}
$n_{\mathrm{pml}}$ & $R_0$ & plateau error & plateau spread \\
\midrule
4  & $10^{-3}$ & $6.7\times10^{-4}$ & $1.0\times10^{-9}$ \\
6  & $10^{-3}$ & $2.6\times10^{-4}$ & $3.3\times10^{-10}$ \\
8  & $10^{-3}$ & $2.6\times10^{-4}$ & $2.2\times10^{-10}$ \\
12 & $10^{-3}$ & $3.4\times10^{-4}$ & $1.7\times10^{-9}$ \\
16 & $10^{-3}$ & $4.2\times10^{-4}$ & $3.3\times10^{-5}$ \\
\midrule
12 & $10^{-2}$ & $4.5\times10^{-3}$ & $1.5\times10^{-10}$ \\
12 & $10^{-3}$ & $3.4\times10^{-4}$ & $1.7\times10^{-9}$ \\
12 & $10^{-4}$ & $4.3\times10^{-5}$ & $1.9\times10^{-7}$ \\
\end{tabular}
\end{ruledtabular}
\end{table}

Table~\ref{appD:tab:reflection} interprets as follows.  The curve is
the sum of two competing contributions.  At the thinnest layer
($n_{\mathrm{pml}}=4$) the discrete transition error dominates: the
measured reflection sits $9.4\times$ above the discrete wall-return
prediction.  For $n_{\mathrm{pml}}\ge8$ the transition error is
negligible and the reflection is set by the wall return, which
\emph{rises toward the design value from below} as the layer widens:
node sampling of the polynomial profile overestimates
$\int\sigma\,dx$ by a factor $1+O(1/n_{\mathrm{pml}})$ (exactly
$\Sigma\sigma h/\!\int\!\sigma = 1.38$, $1.25$, $1.19$, $1.13$,
$1.09$ for $n_{\mathrm{pml}}=4$--$16$), so the effective round-trip
reflection is
$R_{\mathrm{eff}} = R_0^{\,\Sigma\sigma h/\!\int\!\sigma} =
7.1\times10^{-5}$, $1.7\times10^{-4}$, $2.7\times10^{-4}$,
$4.2\times10^{-4}$, $5.2\times10^{-4}$ across the sweep; the measured
values track this prediction within $5$--$20\%$ for
$n_{\mathrm{pml}}\ge8$.  Thin layers therefore \emph{over}-perform
the nominal $R_0$ (at the price of larger transition error), and
widening the layer converges the reflection up to the design value.
We verified this directly in the {\footnotesize SEISMIC\_CPML}
reference implementation~\cite{komatitsch2007unsplit} (the 2D
isotropic elastic code, unmodified kernels, same gold-standard
protocol, $\alpha=0$, $K=1$): its $P$-wave reflection is likewise
non-monotonic and rises from $n_{\mathrm{pml}}=12$ to $16$
($5.8\times10^{-4}\to6.5\times10^{-4}$, against a predicted
$6.5\times10^{-4}\to7.2\times10^{-4}$), converging up toward the
design $R_0$ and matching its own discrete-sum prediction within
${\sim}10\%$ for $n_{\mathrm{pml}}\ge12$.  Two scheme-dependent
differences: the staggered elastic scheme samples the profile at both
full and half points, so its quadrature ratio is
$1+3/(4n_{\mathrm{pml}})$ rather than our
${\sim}1+3/(2n_{\mathrm{pml}})$; and the minimum sits at
$n_{\mathrm{pml}}\approx12$ there because $S$-wave transition
reflection (whose design value $R_0^{c_p/c_s}$ is far smaller) and the
coarser wavelength resolution keep the transition term dominant
further out.  The effect is invisible in normal use of that code, where
the layer width is fixed at ${\sim}10$ cells.

The miscalibration admits a one-line remedy, which we provide as an
option (\texttt{calibration='discrete'}): rescale the staggered profile
\emph{pair} by a single common factor so that the characteristic
damping sum $\Sigma\,(\sigma_v+\sigma_w)h/2$ equals the design
integral exactly.  The realised reflection is then flat at the design
value, $0.81$--$1.08\,R_0$ across the whole sweep
$n_{\mathrm{pml}}=4$--$16$ [red squares in
Fig.~\ref{fig:convergence}(a)], i.e.\ the Collino--Tsogka formula
becomes exact on the grid.  The common factor is essential: rescaling
each profile \emph{separately} gives $\sigma_v$ and $\sigma_w$
different damping functions (the node and midpoint quadratures differ),
an in-layer impedance mismatch that reflects at
$O(1/n_{\mathrm{pml}})$---measured at $1.2\times10^{-2}$ to
$6.9\times10^{-2}$, i.e.\ ten to seventy times \emph{worse} than no
calibration at all.  This remedy applies verbatim to classical CPML
codes.  That the level is
$R_0$-controlled rather than an absolute discretisation floor is
established by the lower block: at $n_{\mathrm{pml}}=12$ the measured
reflection tracks the design value over two decades, with ratios
$E/R_0=0.45$, $0.34$, $0.43$---the Collino--Tsogka formula is
predictive within a factor ${\sim}2.3$ in this discretisation, and
lowering $R_0$ lowers the true reflection commensurately, down to
$4.3\times10^{-5}$ at $R_0=10^{-4}$.  This is the floor behaviour
reported for classical CPML implementations by Komatitsch and
Martin~\cite{komatitsch2007unsplit}.  The plateau is flat to
$\le2\times10^{-9}$ (relative) in six of the eight rows; even the two
most weakly damped rows ($n_{\mathrm{pml}}=16$ and $R_0=10^{-4}$) vary
by only $3.3\times10^{-5}$ and $1.9\times10^{-7}$ of their mean across
$T\in\{100,120,140\}$, i.e.\ the round-trip transient has fully decayed and
the quoted numbers are converged in $T$.

\subsection{Recovery order in the \texorpdfstring{$p$}{p} register}
\label{appD:porder}

\begin{table}
\caption{Schr\"odingerisation recovery error versus $p$-register size
$n_p$ for the kinked $e^{-|p|}$ profile and the $C^{1}$ cubic profile of
Ref.~\cite{jin2024recovery} [1D CPML benchmark, $N=32$,
$n_{\mathrm{pml}}=8$, $\sigma_{\max}=1$, $T=30$, $p\in[-18,18)$,
$\Delta p=36/2^{n_p}$].  Errors are
$\lVert z_{\mathrm{rec}}-e^{AT}z_0\rVert_2$ with
$\lVert z_0\rVert_2=1$, recovered at the library-default slice
$p^{*}=3\Delta p$ (valid at any $p^{*}>0$ since $H_1\preceq0$ here).
The fitted orders are least-squares slopes of $\log_2 E$ versus $n_p$.}
\label{appD:tab:profiles}
\begin{ruledtabular}
\begin{tabular}{llll}
$n_p$ & $\Delta p$ & $e^{-|p|}$ profile & $C^{1}$ cubic profile \\
\midrule
6  & 0.563  & $4.1\times10^{-2}$ & $1.3\times10^{-2}$ \\
7  & 0.281  & $1.6\times10^{-2}$ & $2.6\times10^{-3}$ \\
8  & 0.141  & $3.6\times10^{-3}$ & $1.7\times10^{-4}$ \\
9  & 0.0703 & $1.5\times10^{-3}$ & $1.3\times10^{-5}$ \\
10 & 0.0352 & $1.1\times10^{-3}$ & $5.9\times10^{-6}$ \\
11 & 0.0176 & $6.3\times10^{-4}$ & $2.4\times10^{-6}$ \\
\midrule
\multicolumn{2}{l}{order in $\Delta p$ (all points)} & 1.23 & 2.64 \\
\multicolumn{2}{l}{order in $\Delta p$ ($n_p=6$--$9$)} & 1.64 & 3.40 \\
\end{tabular}
\end{ruledtabular}
\end{table}

Table~\ref{appD:tab:profiles} gives the recovery error of the full
Schr\"odingerised pipeline (exact per-mode evolution, no Trotter) as a
function of the $p$-register size, for the two warping profiles of
Sec.~\ref{sec:schro}.  The errors are absolute per unit initial norm;
by $T=30$ the layer has absorbed all but
$\lVert e^{AT}z_0\rVert_2=1.4\times10^{-3}$ of the state, so resolving
the residual field demands the higher-order profile.  Both profiles
meet or exceed their nominal orders [$O(\Delta p)$ kinked,
$O(\Delta p^{2})$ cubic, Remark~4.9 of Ref.~\cite{jin2024recovery}]:
the full-sweep fits give orders $1.23$ and $2.64$, and the pre-floor
fits ($n_p=6$--$9$) give $1.64$ and $3.40$.  Because the default slice
$p^{*}=3\Delta p$ moves with refinement, we re-evaluated both series at
the fixed slice $p^{*}=1$: the fitted orders become $1.64$ and $3.25$,
so the profile ordering is not a slice artefact.  At $n_p=11$ the cubic
profile is ${\sim}260\times$ more accurate ($2.4\times10^{-6}$ versus
$6.3\times10^{-4}$); the flattening of its last two points marks the
approach to the recovery floor of this configuration.

\subsection{Trotter orders at gate level}
\label{appD:trotter}

\begin{table}
\caption{Gate-level Trotter error versus step count [1D CPML, $N=8$,
$n_{\mathrm{pml}}=2$, $\sigma_{\max}=1$, $T=4$, $n_p=5$, $p_{\max}=8$;
transpiled circuits on a statevector simulator; errors
$\lVert z_{\mathrm{circ}}-z_{\mathrm{ref}}\rVert_2$ with
$\lVert z_0\rVert_2=1$].  ``vs exact S.''\ uses the exact
Schr\"odingerised evolution (exact per-mode propagation at the same $p$
discretisation) as reference, isolating the splitting error; ``vs
$e^{AT}z_0$'' uses the full matrix-exponential reference and therefore
contains the $p$-discretisation floor, measured independently as
$1.29\times10^{-2}$.  Order~2 symmetrises both Trotter levels (Strang
outer split, palindromic term order inner).}
\label{appD:tab:trotter}
\begin{ruledtabular}
\begin{tabular}{lllll}
 & \multicolumn{2}{c}{order 1} & \multicolumn{2}{c}{order 2} \\
steps & vs exact S. & vs $e^{AT}z_0$ & vs exact S. & vs $e^{AT}z_0$ \\
\midrule
10 & $1.36\times10^{-1}$ & $1.32\times10^{-1}$ & $3.24\times10^{-2}$ & $3.40\times10^{-2}$ \\
20 & $6.55\times10^{-2}$ & $6.27\times10^{-2}$ & $8.02\times10^{-3}$ & $1.46\times10^{-2}$ \\
40 & $3.25\times10^{-2}$ & $3.11\times10^{-2}$ & $2.00\times10^{-3}$ & $1.29\times10^{-2}$ \\
80 & $1.62\times10^{-2}$ & $1.75\times10^{-2}$ & $4.99\times10^{-4}$ & $1.29\times10^{-2}$ \\
\midrule
slope & $-1.02$ & $-0.98$ & $-2.01$ & $-0.44$\footnotemark[1] \\
\end{tabular}
\end{ruledtabular}
\footnotetext[1]{Floor-limited: not a meaningful order, see text.}
\end{table}

Table~\ref{appD:tab:trotter} separates the two error sources of the
circuit pipeline of Sec.~\ref{sec:circuit}.  Against the exact
Schr\"odingerised evolution the splitting error is clean: fitted slope
$-1.02$ at order~1 and $-2.01$ at order~2 with both Trotter levels
symmetrised [Fig.~\ref{fig:convergence}(d)].  Against the full
$e^{AT}z_0$ reference the order-2 series saturates at
$1.29\times10^{-2}$ from $40$ steps onward---exactly the independently
measured $p$-discretisation floor of this deliberately coarse $n_p=5$
register---so the nominal slope $-0.44$ in that column is
floor-limited rather than a defect of the splitting.  The order-1
series does not saturate within the sweep because its splitting error
stays above the floor.  Enlarging the $p$ register lowers the floor at
the rates of Table~\ref{appD:tab:profiles}.

\subsection{Stagger ablation}
\label{appD:stagger}

The damping profiles are sampled at $x_j=(j+s)h$ with the stagger $s$
matched to where each field lives (Sec.~\ref{sec:cpml}): for the
forward-difference convention, $v$ at integer nodes ($s=0$) and $w$ at
the half-cells $x_{j-1/2}$ ($s=-1/2$).  Rerunning the gold-standard
protocol ($n_{\mathrm{pml}}=12$, $R_0=10^{-3}$) with the $\sigma_w$
stagger ablated gives Table~\ref{appD:tab:stagger}: misplacing the
profile by half a cell costs two orders of magnitude in true
reflection.  The stagger is load-bearing, not cosmetic.

\begin{table}
\caption{Stagger ablation under the gold-standard protocol
(Sec.~\ref{appD:protocol}; 1D CPML, $n_{\mathrm{pml}}=12$,
$R_0=10^{-3}$): plateau reflection versus the stagger offset $s$ of the
$\sigma_w$ sampling, $x_j=(j+s)h$.}
\label{appD:tab:stagger}
\begin{ruledtabular}
\begin{tabular}{lll}
$\sigma_w$ stagger $s$ & plateau error & degradation \\
\midrule
$-1/2$ (correct half-cell) & $3.4\times10^{-4}$ & $1$ \\
$0$ (unstaggered)          & $4.1\times10^{-2}$ & ${\approx}120$ \\
$+1/2$ (wrong sign)        & $8.1\times10^{-2}$ & ${\approx}235$ \\
\end{tabular}
\end{ruledtabular}
\end{table}

\subsection{Test suite}
\label{appD:tests}

The classical and circuit layers are pinned by $24$ unit tests
(\texttt{tests/test\_absorbing.py}), grouped by the property classes
they assert.  (i)~\emph{Generator structure and spectra}: the
Collino--Tsogka profile amplitude and support;
$\max\operatorname{Re}\operatorname{eig}(A)\le0$ to round-off for the
collapsed 1D, general CFS ($\kappa_{\max}=2$, $\alpha_{\max}=0.05$),
and 2D memory-form generators; the Hermitian split $A=H_1+iH_2$ with
$H_1=-\diag(\sigma)\preceq0$ for the collapsed 1D form and the sponge.
(ii)~\emph{Collapse}: the 1D memory form reproduces the collapsed
diagonal damping to $10^{-10}$.  (iii)~\emph{Absorption and reflection
bounds}: interior energy below $10^{-5}$ where the hard-wall reference
retains more than $0.5$; round-trip amplitude reflection below
$5\times10^{-3}$; 2D CPML and sponge energy decay; and the
gold-standard true reflection below $10^{-3}$ on the reduced instance.
(iv)~\emph{Recovery thresholds}: scalar decay recovered to $10^{-4}$;
the cubic profile beating the kinked profile at equal resolution; 1D
recovery below $2\times10^{-3}$ with a plateau-flatness diagnostic; 2D
sponge recovery below $10^{-3}$ with $\lambda^{+}=0$ asserted; and the
2D CPML threshold of Sec.~\ref{sec:lemma} verified in both
directions---accurate above $p^{*}=\lambda^{+}T+1$ and at least ten
times worse below.  (v)~\emph{Circuit-versus-matrix agreement}: the
exact-Hamiltonian-gate path against the numpy per-mode pipeline to
$10^{-8}$ (pinning the QFT/FFT conventions and qubit ordering); the
Trotterised Bell circuit to $5\times10^{-2}$; the separated 2D operator
strings against the block Hamiltonian to $10^{-13}$; the per-step
orders $O(\delta t^{2})$ and $O(\delta t^{3})$ of the projector-aware
Bell circuit and its Strang variant; and the 17-qubit end-to-end
gate-level run to $3\times10^{-2}$.  (vi)~\emph{Guards and
regression}: rejection of out-of-range recovery slices, degenerate
layers, and inconsistent damping vectors; and a regression test pinning
the untouched closed-domain library (the mixed Dirichlet/Neumann
Laplacian stencil).  Every bound above is asserted numerically, so any
change that degrades a spectrum, a reflection, a recovery threshold, or
a circuit convention fails the suite.

%% file: sections/appE.tex
%

\makeatletter
\@ifundefined{proposition}{\newtheorem{proposition}{Proposition}}{}
\makeatother

\section{The Lyapunov symmetrizer: construction, exact properties, and
compressibility data}
\label{appE:symmetrizer}

This appendix collects the classical construction behind
Sec.~\ref{sec:symmetrizer}, the exact identity underlying
Eq.~\eqref{eq:identity} with its proof, the complete conditioning data
of which Table~\ref{tab:kappa} is an excerpt, the grid-size saturation
measurements, and the compressibility study of the symmetrizer $S$.

\subsection{Construction and classical cost}
\label{appE:construction}

The absorbing generator $A$ is marginally stable---we measure
$\max\operatorname{Re}\operatorname{eig}(A)=0$ to round-off
(Sec.~\ref{sec:2d})---so the Lyapunov equation for $A$ itself is
singular.  We therefore shift, $A_\epsilon = A-\epsilon I$, which is
Hurwitz with margin $\epsilon$, and solve
\begin{equation}
  A_\epsilon^{\dagger}W + WA_\epsilon = -I,
  \qquad
  W = \int_0^{\infty} e^{tA_\epsilon^{\dagger}}\,e^{tA_\epsilon}\,dt
  \;\succ\;0,
  \label{appE:eq:lyap}
\end{equation}
the integral being the unique Hermitian solution.  The symmetrizer is
the Hermitian square root $S=W^{1/2}$, with conditioning
$\kappa_2(S)=\sqrt{\lamax(W)/\lambda_{\min}(W)}$.  The shift is undone
classically by the scalar factor $e^{\epsilon T}$
[Eq.~\eqref{eq:lyapunov}].

Numerically, Eq.~\eqref{appE:eq:lyap} is solved by the
Bartels--Stewart algorithm (one complex Schur decomposition of
$A_\epsilon^{\dagger}$ plus a triangular substitution;
\texttt{scipy.linalg.solve\_continuous\_lyapunov}), and $S$, $S^{-1}$
follow from one Hermitian eigendecomposition of $W$---three dense
$O(n^{3})$ kernels in total.  Measured wall times at the largest
completed size, $n=4096$ ($32\times16$ grid,
$\sigma_{\max}=0.5$): the Lyapunov solve takes $268\,$s at
$\epsilon=10^{-2}$ and $244\,$s at $\epsilon=10^{-3}$, the
eigendecomposition of $W$ takes $31\,$s, the residual verification
$2.5$--$2.6\,$s, and the sparse Lanczos evaluation of
$\lambda^{+}=\lamax^{+}(H_1)$ is negligible ($0.05\,$s).  The $n=8192$
point was gated off by cubic extrapolation of the measured $n=4096$
time (projected $40.3$ min per $\epsilon$, over the $25$-min budget)
and by memory (a dense complex $8192^{2}$ matrix occupies $1$ GiB and
the solve holds six to eight of them, against $5.1$ GiB free).  Across
all 18 configurations of Table~\ref{appE:tab:kappa18} the relative
Lyapunov residual
$\lVert A_\epsilon^{\dagger}W+WA_\epsilon+I\rVert_F/\lVert W\rVert_F$
is at most $1.4\times10^{-14}$.

\subsection{An exact identity for the transformed Hermitian part}
\label{appE:exact}

The equality sign in Eq.~\eqref{eq:identity} is not an estimate.  With
the unit right-hand side $-I$ in Eq.~\eqref{appE:eq:lyap}, the
Hermitian part of the transformed generator is an explicit function of
$W$:

\begin{proposition}
\label{appE:prop}
Let $A\in\mathbb{C}^{n\times n}$ have spectral abscissa at most zero,
fix $\epsilon>0$, set $A_\epsilon=A-\epsilon I$, let $W\succ0$ solve
$A_\epsilon^{\dagger}W+WA_\epsilon=-I$, and let $S=W^{1/2}$ be the
Hermitian positive-definite square root.  Then, exactly,
\begin{align}
  \tfrac12\bigl(SA_\epsilon S^{-1}+S^{-1}A_\epsilon^{\dagger}S\bigr)
    &= -\tfrac12 W^{-1} \;\prec\; 0,
  \label{appE:eq:identshift}\\
  \tfrac12\bigl(SAS^{-1}+S^{-1}A^{\dagger}S\bigr)
    &= \epsilon I - \tfrac12 W^{-1}.
  \label{appE:eq:identorig}
\end{align}
In particular the $\epsilon$-shifted transform is strictly dissipative
with $\lamax(\Ht^{(\epsilon)}) = -1/\bigl(2\lamax(W)\bigr)$, and the
transform of the original $A$ has
$\lamax(\Ht) = \epsilon - 1/\bigl(2\lamax(W)\bigr) < \epsilon$.
\end{proposition}

\begin{lemproof}
Since $A_\epsilon$ is Hurwitz, the integral in
Eq.~\eqref{appE:eq:lyap} converges and is the unique Hermitian
solution; it is positive definite, so $S=W^{1/2}$ exists, is
Hermitian, and $S^{2}=W$.  Write
$\At=SA_\epsilon S^{-1}$ and take any $x\in\mathbb{C}^{n}$ with
$x=Sy$.  By congruence,
\begin{equation}
  x^{\dagger}\bigl(\At+\At^{\dagger}\bigr)x
  = y^{\dagger}\bigl(WA_\epsilon+A_\epsilon^{\dagger}W\bigr)y
  = -\lVert y\rVert^{2}
  = -x^{\dagger}W^{-1}x,
  \label{appE:eq:congruence}
\end{equation}
using $S\At S = WA_\epsilon$ and $S\At^{\dagger}S =
A_\epsilon^{\dagger}W$.  Two Hermitian matrices with equal quadratic
forms are equal, so $\At+\At^{\dagger}=-W^{-1}$, which is
Eq.~\eqref{appE:eq:identshift}; its largest eigenvalue is
$-\tfrac12\lambda_{\min}(W^{-1})=-1/(2\lamax(W))$.  For
Eq.~\eqref{appE:eq:identorig}, $SAS^{-1}=SA_\epsilon
S^{-1}+\epsilon I$ and $\epsilon I$ is Hermitian, so the Hermitian
part shifts by $\epsilon I$ and the eigenvalues are
$\epsilon-1/(2\lambda_j(W))$.
\end{lemproof}

The identity is verified in every measured configuration: in all 18
rows of Table~\ref{appE:tab:kappa18} the directly computed
$\lamax(\Ht)$ (dense eigendecomposition of the Hermitian part of
$SAS^{-1}$) agrees with the prediction $\epsilon-1/(2\lamax(W))$ to a
relative deviation of at most $3.8\times10^{-12}$---far beyond the
four-digit agreement quoted in the main text---and the shifted column
equals the unshifted one minus $\epsilon$ to the same precision.  A
direct check of the matrix identity~\eqref{appE:eq:identshift} on an
$n=50$ instance gives a Frobenius residual of $8.9\times10^{-14}$.
Proposition~\ref{appE:prop} also explains the choice made in
Sec.~\ref{sec:symmetrizer}: the transform of the \emph{original} $A$
is not dissipative ($\lamax(\Ht)\approx\epsilon>0$), whereas the
transform of $A_\epsilon$ is, strictly; we therefore evolve the
shifted system and undo $e^{\epsilon T}$ classically.

\begin{table*}
\caption{Full Lyapunov-symmetrizer data set; Table~\ref{tab:kappa} of
the main text is the excerpt obtained by dropping the
$\lambda_{\min}(W)$ and unshifted-transform columns.  Grids are listed
in points ($n$ is the full state dimension including the eight
blocks).  $\lambda^{+}=\lamax^{+}(H_1)$ is the indefiniteness of the
untransformed generator; $\kappa_2(S)=\sqrt{\operatorname{cond}_2(W)}$;
$\lamax(\Ht)$ is the largest Hermitian-part eigenvalue of the
transform $SAS^{-1}$ of the original generator and
$\lamax(\Ht^{(\epsilon)})$ that of the $\epsilon$-shifted transform
$SA_\epsilon S^{-1}$; $T^{*}=\ln\kappa^{2}/(2\lambda^{+})$ is the
crossover time.  Every row satisfies the identities of
Proposition~\ref{appE:prop}: $\lamax(\Ht)$ equals
$\epsilon-1/(2\lamax(W))$ to a relative deviation
$\le3.8\times10^{-12}$, and
$\lamax(\Ht^{(\epsilon)})=\lamax(\Ht)-\epsilon$.  Relative Lyapunov
residuals are $\le1.4\times10^{-14}$.}
\label{appE:tab:kappa18}
\begin{ruledtabular}
\begin{tabular}{llllllllll}
Grid & $n$ & $\sigma_{\max}$ & $\epsilon$ & $\lambda^{+}(H_1)$ &
$\lambda_{\min}(W)$ & $\kappa_2(S)$ & $\lamax(\Ht)$ &
$\lamax(\Ht^{(\epsilon)})$ & $T^{*}$ \\
\colrule
$8\times8$ & 512 & 0.5 & $10^{-2}$ & 0.683 & 0.669 & $402.1$ & $9.9954\times10^{-3}$ & $-4.63\times10^{-6}$ & 8.8 \\
$8\times8$ & 512 & 0.5 & $10^{-3}$ & 0.683 & 0.681 & $3009$ & $9.9992\times10^{-4}$ & $-8.10\times10^{-8}$ & 11.7 \\
$8\times8$ & 512 & 0.5 & $10^{-4}$ & 0.683 & 0.683 & $1.105\times10^{4}$ & $9.9994\times10^{-5}$ & $-6.00\times10^{-9}$ & 13.6 \\
$8\times8$ & 512 & 1.0 & $10^{-2}$ & 0.927 & 0.373 & $509.3$ & $9.9948\times10^{-3}$ & $-5.17\times10^{-6}$ & 6.7 \\
$8\times8$ & 512 & 1.0 & $10^{-3}$ & 0.927 & 0.376 & $2851$ & $9.9984\times10^{-4}$ & $-1.64\times10^{-7}$ & 8.6 \\
$8\times8$ & 512 & 1.0 & $10^{-4}$ & 0.927 & 0.377 & $9762$ & $9.9986\times10^{-5}$ & $-1.39\times10^{-8}$ & 9.9 \\
$16\times8$ & 1024 & 0.5 & $10^{-2}$ & 0.687 & 0.661 & $506.3$ & $9.9970\times10^{-3}$ & $-2.95\times10^{-6}$ & 9.1 \\
$16\times8$ & 1024 & 0.5 & $10^{-3}$ & 0.687 & 0.673 & $7009$ & $9.9998\times10^{-4}$ & $-1.51\times10^{-8}$ & 12.9 \\
$16\times8$ & 1024 & 0.5 & $10^{-4}$ & 0.687 & 0.674 & $3.052\times10^{4}$ & $9.9999\times10^{-5}$ & $-7.96\times10^{-10}$ & 15.0 \\
$16\times8$ & 1024 & 1.0 & $10^{-2}$ & 0.956 & 0.369 & $749.6$ & $9.9976\times10^{-3}$ & $-2.41\times10^{-6}$ & 6.9 \\
$16\times8$ & 1024 & 1.0 & $10^{-3}$ & 0.956 & 0.372 & $6981$ & $9.9997\times10^{-4}$ & $-2.76\times10^{-8}$ & 9.3 \\
$16\times8$ & 1024 & 1.0 & $10^{-4}$ & 0.956 & 0.373 & $2.732\times10^{4}$ & $9.9998\times10^{-5}$ & $-1.80\times10^{-9}$ & 10.7 \\
$16\times16$ & 2048 & 0.5 & $10^{-2}$ & 0.694 & 0.655 & $566.9$ & $9.9976\times10^{-3}$ & $-2.37\times10^{-6}$ & 9.1 \\
$16\times16$ & 2048 & 0.5 & $10^{-3}$ & 0.694 & 0.667 & $9856$ & $9.9999\times10^{-4}$ & $-7.71\times10^{-9}$ & 13.3 \\
$16\times16$ & 2048 & 0.5 & $10^{-4}$ & 0.694 & 0.669 & $6.612\times10^{4}$ & $1.0000\times10^{-4}$ & $-1.71\times10^{-10}$ & 16.0 \\
$16\times16$ & 2048 & 1.0 & $10^{-2}$ & 0.972 & 0.367 & $857.7$ & $9.9981\times10^{-3}$ & $-1.85\times10^{-6}$ & 6.9 \\
$16\times16$ & 2048 & 1.0 & $10^{-3}$ & 0.972 & 0.370 & $1.249\times10^{4}$ & $9.9999\times10^{-4}$ & $-8.67\times10^{-9}$ & 9.7 \\
$16\times16$ & 2048 & 1.0 & $10^{-4}$ & 0.972 & 0.370 & $6.347\times10^{4}$ & $1.0000\times10^{-4}$ & $-3.35\times10^{-10}$ & 11.4 \\
\end{tabular}
\end{ruledtabular}
\end{table*}

\subsection{The full conditioning table}
\label{appE:fulltable}

Table~\ref{appE:tab:kappa18} lists all 18 measured configurations
(three grids, two damping strengths, three shifts).  Three regularities
are worth recording.  (i)~$\lambda_{\min}(W)$ is essentially
independent of $\epsilon$ ($0.66$--$0.68$ at $\sigma_{\max}=0.5$,
$0.37$ at $\sigma_{\max}=1.0$), so the entire $\epsilon$ dependence of
$\kappa_2(S)$ is carried by $\lamax(W)$, consistent with
Proposition~\ref{appE:prop}, in which $\lamax(W)$ is pinned to the
inverse dissipativity margin of the shifted transform.
(ii)~Least-squares fits of $\ln\kappa_2(S)$ against $\ln\epsilon$ per
(grid, $\sigma_{\max}$) configuration give slopes between $-0.64$ and
$-1.03$ (the range quoted as
$\epsilon^{-0.6\,\text{to}\,-1.0}$ in Sec.~\ref{sec:symmetrizer}),
steepening with grid size at fixed $\epsilon$ range.  (iii)~The
crossover times $T^{*}=\ln\kappa^{2}/(2\lambda^{+})$ span
$6.7$--$16.0$ and grow only logarithmically with $\kappa_2(S)$.  For
comparison, the eigenvector symmetrizer $W=(VV^{\dagger})^{-1}$
yields $\kappa_2(S)=2.9\times10^{8}$--$6.9\times10^{8}$ where it
exists and fails outright ($VV^{\dagger}$ numerically singular) on
both $16\times8$ configurations; the Lyapunov construction is four to
six orders of magnitude better conditioned and never fails.

\subsection{Saturation of \texorpdfstring{$\kappa_2(S)$}{kappa2(S)}
with grid size}
\label{appE:saturation}

Table~\ref{appE:tab:scaling} extends the $\sigma_{\max}=0.5$ series to
$n=4096$.  At $\epsilon=10^{-2}$ the conditioning is flat:
$\kappa_2(S)$ \emph{decreases} from $566.9$ to $564.5$ between
$n=2048$ and $n=4096$ (local log--log slope $-0.006$).  At
$\epsilon=10^{-3}$ it still rises but visibly flattens.  Power-law
fits confirm that a power law is the wrong model: at
$\epsilon=10^{-2}$ the fitted exponent drops from $0.25$
($r^{2}=0.963$, $n=512$--$2048$) to $0.16$ ($r^{2}=0.823$,
$n=512$--$4096$), and at $\epsilon=10^{-3}$ from $0.86$
($r^{2}=0.943$) to $0.62$ ($r^{2}=0.875$)---falling exponents with
degrading fit quality, i.e.\ $\kappa$ is saturating, not following a
power law.  The drivers saturate individually: $\lambda^{+}$ grows
from $0.6829$ to $0.6940$ over the eightfold range in $n$ and is then
identical to ten digits between $n=2048$ and $n=4096$, and
$\lambda_{\min}(W)$ stays at $0.655$--$0.681$ throughout.  The
conditioning is therefore $\epsilon$-controlled, not grid-controlled:
mesh refinement does not inflate the symmetrizer cost.

\begin{table}
\caption{$\kappa_2(S)$ and $\lambda_{\min}(W)$ versus grid size at
$\sigma_{\max}=0.5$ (grids in points; $n$ is the state dimension).
The $n=8192$ point was gated off by the projected runtime and memory
(Sec.~\ref{appE:construction}).}
\label{appE:tab:scaling}
\begin{ruledtabular}
\begin{tabular}{lllllll}
 & & & \multicolumn{2}{c}{$\epsilon=10^{-2}$} &
 \multicolumn{2}{c}{$\epsilon=10^{-3}$} \\
\cline{4-5}\cline{6-7}
Grid & $n$ & $\lambda^{+}(H_1)$ & $\kappa_2(S)$ & $\lambda_{\min}(W)$
 & $\kappa_2(S)$ & $\lambda_{\min}(W)$ \\
\colrule
$8\times8$ & 512 & 0.6829 & 402.1 & 0.669 & 3009 & 0.681 \\
$16\times8$ & 1024 & 0.6875 & 506.3 & 0.661 & 7009 & 0.673 \\
$16\times16$ & 2048 & 0.6940 & 566.9 & 0.655 & 9856 & 0.667 \\
$32\times16$ & 4096 & 0.6940 & 564.5 & 0.655 & 11241 & 0.668 \\
\end{tabular}
\end{ruledtabular}
\end{table}

\subsection{Compressibility of \texorpdfstring{$S$}{S}}
\label{appE:compress}

\paragraph{Protocol.}  At $\epsilon=10^{-2}$, $\sigma_{\max}=0.5$, we
measured the structure of $S$ on the $n=512$ grid and confirmed on the
$n=1024$ grid, and scored 25 candidate approximate symmetrizers
$S_b\approx S$ drawn from six families (Manhattan-distance-banded,
magnitude-thresholded, block-pattern, identity-plus-low-rank,
low-rank-plus-band hybrid, and banded-$S^{-1}$).  Each candidate is
scored by its residual indefiniteness
$\lambda_b = \lamax\bigl(\mathrm{Herm}(S_bA_\epsilon
S_b^{-1})\bigr)$; for the exact $S$,
$\lambda_b=-1/(2\lamax(W))<0$ by Proposition~\ref{appE:prop}.  A
candidate \emph{passes} if $\lambda_b\le2\times10^{-3}$ (the
$\epsilon$-mode recovery at the default slice remains valid); a
candidate with $2\times10^{-3}<\lambda_b\le0.5$ is
\emph{rescue-eligible}: the classical shift is enlarged from
$\epsilon$ to $s=\epsilon+\lambda_b$, so the evolved generator
$S_b(A-sI)S_b^{-1}$ is again dissipative, the undo factor becomes
$e^{sT}$, and the post-selection penalty relative to the exact $S$ has
exponent $2\lambda_bT$.  End-to-end recoveries (errors relative to
$e^{AT}z_0$ at $T\in\{10,40\}$, $n_p=10$) were run for the exact $S$
and one representative per surviving family.
Table~\ref{appE:tab:compress} lists all candidates on the $n=1024$
confirmation grid.

\paragraph{Structure of $S$.}  Three exact or near-exact structural
facts.  (i)~\emph{Dead-index identity.}  On the $328$ of $1024$
indices ($32\%$; $176$ of $512$, i.e.\ $34\%$, at $n=512$) that carry
no dynamics---interior memory components and the zero padding
block---$A$ acts as zero, so Eq.~\eqref{appE:eq:lyap} reduces to
$2\epsilon w=1$ there and $S$ is \emph{exactly}
$(2\epsilon)^{-1/2}I$ on this subspace: measured deviation
$2.0\times10^{-14}$ on the diagonal, off-diagonal entries below
$7.8\times10^{-15}$, and coupling to the active subspace below
$1.1\times10^{-12}$.  This factor is free structure for any
implementation.  (ii)~\emph{The active support is about half the
matrix}: the exact $S$ has $48.1\%$ ($n=1024$) and $51.1\%$
($n=512$) nonzero entries (at threshold $10^{-14}$ of the largest).
(iii)~\emph{The correction is full rank but strip-localised}: with the
optimal scalar $\alpha=16.7$, the correction $E=S-\alpha I$ carries
$94.5\%$ of the Frobenius norm of $S$, has $999$ of $1024$ singular
values above $10^{-2}$ of the largest, and needs rank $995$ to capture
$99.99\%$ of its energy---numerically full rank---while $99.6\%$ of
that energy sits on index pairs both lying in the absorbing-layer
strips ($72$ of the $1024$ indices are strip-free interior).

\begin{table*}
\caption{All 25 candidate approximate symmetrizers on the $n=1024$
($16\times8$) confirmation grid at $\epsilon=10^{-2}$,
$\sigma_{\max}=0.5$.  Density is the fraction of nonzero entries of
the sparse object; for the $\alpha I+{}$rank-$r$ family it is
$(r{+}1)/n$ (the number of LCU terms relative to $n$), and for the
banded-$S^{-1}$ family it is the density of the banded $S^{-1}$ (state
preparation by sparse linear solve).
$\lambda_b=\lamax\bigl(\mathrm{Herm}(S_bA_\epsilon S_b^{-1})\bigr)$ is
the residual indefiniteness.  Status: ``pass'' $=$
$\lambda_b\le2\times10^{-3}$, valid in $\epsilon$ mode at the default
slice; ``rescue'' $=$ run end to end with the enlarged shift
$\epsilon+\lambda_b$; ``elig.''\ $=$ rescue-eligible
($\lambda_b\le0.5$) but not run end to end; ``fail'' $=$ $S_b$
indefinite or $\lambda_b>0.5$.  The top-$50\%$ threshold and the
support-of-$A^{2}$ patterns reproduce the exact $S$ to machine
precision (relative Frobenius deviation $\le6\times10^{-16}$); the
end-to-end errors of every usable candidate are obtained at
conditioning $\operatorname{cond}(S_b)=5.06\times10^{2}$--$5.11
\times10^{2}$, equal to the exact value $5.06\times10^{2}$.}
\label{appE:tab:compress}
\begin{ruledtabular}
\begin{tabular}{llllll}
Candidate $S_b$ & density (\%) & $\lambda_b$ & status &
err.\ ($T{=}10$) & err.\ ($T{=}40$) \\
\colrule
exact $S$ (active support) & 48.1 & $-3.0\times10^{-6}$ & pass & $7.7\times10^{-6}$ & $4.9\times10^{-5}$ \\
Manhattan band, $d=1$ & 1.8 & $2.36$ & fail & -- & -- \\
Manhattan band, $d=2$ & 4.2 & $1.32$ & fail & -- & -- \\
Manhattan band, $d=4$ & 10.8 & $3.09$ & fail & -- & -- \\
Manhattan band, $d=8$ & 26.5 & $0.062$ & elig. & -- & -- \\
Manhattan band, $d=12$ & 38.6 & $0.056$ & rescue & $8.7\times10^{-6}$ & $3.1\times10^{-4}$ \\
magnitude threshold, top 1\% & 1.0 & $104$ & fail & -- & -- \\
magnitude threshold, top 5\% & 5.0 & $3.03$ & fail & -- & -- \\
magnitude threshold, top 10\% & 10.0 & $0.734$ & fail & -- & -- \\
magnitude threshold, top 25\% & 25.0 & $0.054$ & elig. & -- & -- \\
magnitude threshold, top 35\% & 35.0 & $4.2\times10^{-3}$ & rescue & $8.7\times10^{-6}$ & $5.4\times10^{-5}$ \\
magnitude threshold, top 50\% & 48.1 & $-3.0\times10^{-6}$ & pass & -- & -- \\
block diagonal (per-field) & 7.4 & $6.83$ & fail & -- & -- \\
support of $A$ & 29.9 & $8.58$ & fail & -- & -- \\
support of $A^{2}$ & 48.1 & $-3.0\times10^{-6}$ & pass & $7.7\times10^{-6}$ & $4.9\times10^{-5}$ \\
$\alpha I+{}$rank 8 & 0.9 & $0.677$ & fail & -- & -- \\
$\alpha I+{}$rank 16 & 1.7 & $0.677$ & fail & -- & -- \\
$\alpha I+{}$rank 32 & 3.2 & $0.674$ & fail & -- & -- \\
$\alpha I+{}$rank 64 & 6.3 & $0.674$ & fail & -- & -- \\
$\alpha I+{}$rank 32${}+{}$band 2 & 4.2 & $0.374$ & elig. & -- & -- \\
$\alpha I+{}$rank 32${}+{}$band 8 & 26.5 & $0.066$ & rescue & $1.3\times10^{-5}$ & $6.6\times10^{-4}$ \\
$\alpha I+{}$rank 64${}+{}$band 2 & 4.2 & $0.349$ & elig. & -- & -- \\
$\alpha I+{}$rank 64${}+{}$band 8 & 26.5 & $0.033$ & elig. & -- & -- \\
banded $S^{-1}$, $d=2$ & 4.0 & $5.02$ & fail & -- & -- \\
banded $S^{-1}$, $d=8$ & 25.3 & $2.23$ & fail & -- & -- \\
\end{tabular}
\end{ruledtabular}
\end{table*}

\paragraph{Findings.}  The full-rank measurement rules out
identity-plus-low-rank outright: for ranks $8$--$64$ the residual
$\lambda_b=0.674$--$0.677$ equals the \emph{untransformed}
$\lambda^{+}=0.687$ to within two percent, i.e.\ a short LCU
correction has no symmetrizing effect at all.  Narrow bands,
aggressive thresholds, block patterns, and banded inverses are either
indefinite or carry $\lambda_b\gtrsim1$.  Moderate compression,
however, works under the rescue accounting: magnitude thresholding at
$35\%$ density recovers to $8.7\times10^{-6}$ ($T=10$) and
$5.4\times10^{-5}$ ($T=40$); the Manhattan band $d=12$ at $38.6\%$
reaches $8.7\times10^{-6}$ and $3.1\times10^{-4}$; the rank-32 plus
band-8 hybrid at $26.5\%$ reaches $1.3\times10^{-5}$ and
$6.6\times10^{-4}$.  The rescue cost is mild at the measured
residuals: the largest rescued $\lambda_b=0.066$ gives an undo factor
$e^{\lambda_bT}=e^{2.6}\approx14$ and a post-selection penalty
$e^{2\lambda_bT}=e^{5.3}\approx2\times10^{2}$ at $T=40$, against the
$T$-independent baseline $\kappa_2^{2}(S)\approx2.6\times10^{5}$.  The
rescue frame is also the robust one: the dissipativity margin of the
exact $S$ is only $|\lambda_b|=3.0\times10^{-6}$, and a perturbation
of $S$ of relative spectral norm $6.4\times10^{-7}$ suffices to push
$\lambda_b$ past the $2\times10^{-3}$ gate, so \emph{strict}
dissipativity of any approximation is fragile, whereas absorbing the
measured $\lambda_b$ into the classical shift is exact bookkeeping.
Taken together---half-dense active support, full-rank strip-localised
correction, free dead-index identity factor---these measurements
support the implementation verdict of Sec.~\ref{sec:symmetrizer}:
block-encoding of a ${\sim}35\%$-dense operator rather than a local
circuit or a short LCU, with a structured-ansatz Lyapunov solve as the
open problem.  The compression study is limited to $n\le1024$.

\subsection{Choosing the shift \texorpdfstring{$\epsilon$}{epsilon}}
\label{appE:eps}

The shift enters the cost twice and the accuracy nowhere: the
recovered state is exact up to the classically applied factor
$e^{\epsilon T}$, while $\kappa_2(S)$ grows as
$\epsilon^{-0.64}$--$\epsilon^{-1.03}$
(Sec.~\ref{appE:fulltable}) and the evolved norm of the shifted system
costs $e^{2\epsilon T}$ in post-selection.  At the recommended
$\epsilon=10^{-2}$ these are $\kappa_2(S)=4.0\times10^{2}$--$8.6
\times10^{2}$ across all configurations and
$e^{\epsilon T}\le e^{0.4}$ ($e^{2\epsilon T}\le e^{0.8}\approx2.2$)
over the horizons tested ($T\le40$)---the shift is negligible against
$\kappa_2^{2}$, and decreasing $\epsilon$ by a decade buys only that
trivial factor while inflating $\kappa_2^{2}$ by one and a half to two
decades.  The shifted transform is what makes the large-$\epsilon$
choice safe: with the \emph{unshifted} transform of $A$,
Proposition~\ref{appE:prop} gives
$\lamax(\Ht)=\epsilon-1/(2\lamax(W))\approx\epsilon>0$, so the
recovery threshold [Eq.~\eqref{eq:recovery}] would demand
$p^{*}\gtrsim\epsilon T$---at the default slice $p^{*}=0.117$
($n_p=10$, $p_{\max}=20$) and $T=40$ this would force
$\epsilon\lesssim2.9\times10^{-3}$, excluding the best-conditioned
setting, and would couple $\epsilon$ to the horizon and the
$p$-register extent.  The shifted transform has
$\Ht^{(\epsilon)}\preceq-1/(2\lamax(W))\,I\prec0$, so any $p^{*}>0$ is
valid at every $T$ and $\epsilon$ acts purely as a conditioning knob.